\documentclass[12pt,reqno,a4paper]{article}
\usepackage{extsizes}
\usepackage{arxiv}
\usepackage{authblk}
\usepackage[T1]{fontenc}
\usepackage{lmodern}   
\usepackage{url}            
\usepackage{booktabs}       
\usepackage{amsfonts}       
\usepackage{nicefrac}       
\usepackage{microtype}      
\usepackage{graphicx}
\usepackage{caption}
\usepackage{subcaption}
\usepackage{float}
\usepackage[sorting=none, style=ieee, citestyle=numeric-comp]{biblatex}
\usepackage{comment}
\usepackage{gensymb} 
\usepackage{siunitx}
\usepackage{upgreek}
\usepackage{amsmath}
\usepackage{amssymb}
\usepackage{graphicx}
\graphicspath{ {./figures/} {./FIG_SUPPLEMENTARY/} }

\renewcommand{\headeright}{}
\renewcommand{\undertitle}{}
\renewcommand{\shorttitle}{\emph{arXiv} Preprint}

\author[1]{Francesco Vitale*}
\author[2]{Stephen A. Church}
\author[3]{Daniel Repp}
\author[4]{Karthika S. Sunil}
\author[4]{Mario Ziegler}
\author[4]{Marco Diegel}
\author[4]{Andrea Dellith}
\author[5]{Thi-Hien Do}
\author[5]{Sheng-Di Lin}
\author[4,6,7,8]{Jer-Shing Huang}
\author[3]{Thomas Pertsch}
\author[2]{Patrick Parkinson}
\author[1]{Carsten Ronning**}

\affil[1]{Institute of Solid State Phyisics, Friedrich Schiller University Jena, Max-Wien-Platz 1, 07743 Jena, Germany}
\affil[2]{Department of Physics and Astronomy and Photon Science Institute, the University of Manchester, Manchester M13 9PL, United Kingdom}
\affil[3]{Institute of Applied Physics, Friedrich Schiller University Jena, Albert-Einstein-Straße 15, 07745 Jena, Germany
}
\affil[4]{Leibniz Institute of Photonic Technology, Albert-Einstein-Straße 9, 07745 Jena, Germany}
\affil[5]{Institute of Electronics National Yang-Ming Chiao-Tung University, Hsinchu 30010, Taiwan}
\affil[6]{Institute of Physical Chemistry and Abbe Center of Photonics, Friedrich Schiller University Jena, Helmholtzweg 4, 07743 Jena, Germany}
\affil[7]{Department of Electrophysics, National Yang-Ming Chiao-Tung University, 1001 Ta-Hsueh Road, Hsinchu 30010, Taiwan}
\affil[8]{Research Center for Applied Sciences, Academia Sinica, 128 Sec. 2 Academia Road, Taipei 11529, Taiwan}

\date{}

\title{Optical Characterization of Size- and Substrate-Dependent Performance of Ultraviolet Hybrid Plasmonic Nanowire Lasers}

\begin{document}
	\maketitle

	*email: francesco.vitale@uni-jena.de \\
	**email: carsten.ronning@uni-jena.de
	
\vspace{1.5cm}

	\begin{abstract}
		Nanowire-based plasmonic lasers are now established as nano-sources of coherent radiation, appearing as suitable candidates for integration into next-generation nanophotonic circuitry. However, compared to their photonic counterparts, their relatively high losses and large lasing thresholds still pose a burdening constraint on their scalability. In this study, the lasing characteristics of ZnO nanowires on Ag and Al substrates, operating as optically-pumped short-wavelength plasmonic nanolasers, are systematically investigated in combination with the size-dependent performance of the hybrid cavity. A hybrid nanomanipulation-assisted single nanowire optical characterization combined with high-throughput PL spectroscopy enables the correlation of the lasing characteristics to the metal substrate and the nanowire diameter. The results evidence that the coupling between excitons and surface plasmons is closely tied to the relationship between substrate dispersive behavior and nanowire diameter. Such coupling dictates the degree to which the lasing character, be it more plasmonic- or photonic-like, can define the stimulated emission features and, as a result, the device performance.

	\end{abstract}
	

	\section{Introduction}
	
	Following the conceptualization of the SPASER (Surface Plasmon Amplification by Stimulated Emission of Radiation) by Bergmann and Stockmann \cite{bergman2003} at the turn of the millennium, the field of plasmonic nanolasers has marked a turning point in the development of sub-wavelength hybrid waveguides of coherent radiation for next-generation all-optical circuitry \cite{ozbay2006, davis2017, fukuda2020}. Within this framework, semiconductor nanowire-based plasmonic nanolasers have gained a foothold since the first experimental realization by Oulton \emph{et al.}, designed as a semiconductor-insulator-metal (SIM) platform \cite{oulton2008,oulton2009} in which lasing is sustained by surface plasmon polaritons (SPPs). Thereby, in contrast to their photonic counterparts \cite{voss2007,zimmler2008}, nanowire-based plasmonic lasers have proven to support lasing modes whose size can surpass the diffraction limit \cite{ma2011,wang2017}, however at the cost of higher lasing thresholds that typically arise from the quasi-direct contact of the semiconductor nanowire with the metal layer \cite{zhang2014,liang2020}. On the other hand, the sub-wavelength confinement of the light field ensures smaller mode volumes and larger Purcell factors \cite{chou2015,wei2016}, leading to an acceleration of the lasing dynamics \cite{sidiropoulos2014,lu2017}. Conversely, photonic nanowire-based lasers show robust device performances, mainly benefitting from lower lasing thresholds and higher optical damage thresholds \cite{zhang2009l,zapf2017,church2022}. Nonetheless, the diameter cutoff for cavity-waveguided modes limits their scalability to $d \approx$ 150 nm below which only hybrid plasmonic modes can efficiently sustain lasing action \cite{saxena2013,roeder2014}. For instance, it has been shown for photonic devices that ZnO nanowires in the diameter range between 150 and 200 nm, exhibit a lasing action predominantly supported by HE$_{11}$ modes \cite{roeder2016}, while above 200 nm, a TE$_{01}$ mode typically dominates the mode competition, with only little contribution from the HE$_{11}$ modes \cite{buschlinger2015}. On the other hand, short-length ($L < 2 \, \unit{\upmu m}$) nanowires with diameters well below the cutoff for waveguided modes, have been reliably employed for sustaining a purely plasmonic lasing action in hybrid cavities coupled with metallic substrates. Despite that, such devices usually suffer from high lasing thresholds and relatively low optical damage thresholds, especially at room temperature \cite{chouYH2016,chou2017}. However, a more complete and detailed understanding of the mode competition for devices featuring nanowires with diameters in the “transition region” between purely plasmonic ($<$ 150 nm) and purely photonic ($>$ 200 nm) nanolasers has not yet been addressed. \\
	
	In this context, ZnO nanowires have been extensively utilized as the building block of plasmonic nanolasers designed for the ultraviolet (UV) range, demonstrating relevance for the envisaged nano-device scalability in the pursuit of low-background, all-optical communication \cite{notomi2020,gu2021}.  Indeed, ZnO has been largely employed as a gain medium for plasmonic platforms, thanks to the high material gain and large exciton binding energies \cite{madelung2004} that lead to a stable electron-hole plasma formation at room temperature \cite{vanmaekelbergh2011}, and also promote surface plasmon– exciton polariton coupling \cite{zhao2020}. Regarding the metallic substrate, aluminum (Al) and silver (Ag) have represented a valuable choice for the realization of short-wavelength plasmonic nanolasers, thanks to their surface plasmon resonance lying in the UV range \cite{lu2012,gwo2016,liu2015,chouBT2016}. On the one hand, silver possesses an interband transition around 360 nm which leads to a characteristic back-bending of the surface plasmon dispersion \cite{chou2015}: therefore, ZnO nanowire-based nanolasers coupled with an Ag substrate are typically operating at frequencies close to the metal surface plasmon frequency occurring near the excitonic resonance of ZnO at $\approx 370 \, \unit{nm}$) \cite{chou2017}. Conversely, aluminum is characterized by an energetically higher surface plasmon resonance (around 200 nm) than silver, lying far from the gain medium excitonic resonance and, thus, indicative of a less favorable plasmonic coupling  \cite{knight2014,chou2017}. Although a fundamental understanding of the plasmonic properties of both metals is quite established, a one-to-one comparison between the two as “plasmonic substrates” along with a more detailed comprehension of the nanowire dimensions affecting the mode competition within the hybrid nano-cavity, remains unexplored. Indeed, the nanowire-to-nanowire inhomogeneity typically makes it challenging to compare the different nanolaser classes robustly  \cite{brewster2012,alabri2021}. Therefore, a substrate-related understanding of the lasing properties from single nanowires has been used in this study, to shed more light on how the different metallic substrates can influence the lasing properties of nanowire-based plasmonic lasers in combination with their cavity dimensions, in an attempt to draw guidelines for developing "mixed" plasmonic-photonic devices incorporating the advantages of both counterparts, in combination with other modification schemes of the hybrid cavity \cite{chou2018,li2019,vitale2023}. In this study, we investigate the impact of Ag and Al substrates on the lasing characteristics of ZnO nanowires by transferring a single nanowire between each substrate and measuring the resulting change in the lasing characteristics. We then demonstrate the generality of our conclusions by scaling up the study using automated, high-throughput photoluminescence to study the lasing action of more than 2,000 nanowires that are randomly distributed across both substrates. These experimental approaches provide a rigorous assessment of the performance of each substrate, compensated for the statistical fluctuations resulting from variations in crystal quality, cavity size (both length and diameter), and morphology (whether tapered, strained, roughened, or cleaved at the end facets) across different wires \cite{couteau2015,zapf2019}, which impacted previous studies.
	
	\section{Results and Discussion}
	
	\subsection{Manipulation-assisted optical investigation of single hybrid nanowire cavities}
	
	In the first part of this study, we employed a micro-photoluminescence ($\upmu$-PL) setup equipped with a nanomanipulator, to investigate the lasing properties of the same ZnO nanowire when moved between the two different metallic substrates (i.e. silver and aluminum) as schematically shown in \textbf{Figure 1}(a). 
	
	The single-nanowire PL measurements targeted individual nanowires dry-imprinted on Ag flakes, as the one exemplarily shown in Figure 1(b), which were then moved onto the Al layer by contact with the nanomanipulator W tip and subsequently re-measured. Both the Ag flakes and the Al layer exhibited a very high crystal quality, as evidenced by the XRD plot in Figure 1(c) showing their corresponding sharp (111) peaks. The full 2$\theta$ scan, along with a more extensive peak indexing, can be found in the Supplementary Information with the corresponding reciprocal space maps. The narrow (111) peak for the Al layer is a result of the high orientation arising from the epitaxial growth, while it is indicative of large grain size for the Ag flakes, which are also characterized by a highly smooth surface, as evidenced by the AFM map in Figure 1(d). The average root-mean-square (RMS) surface roughness was found to be $R_q = 0.9 \pm 0.2 \, \unit{nm}$ for the Ag flakes, and $R_q = 1.01 \pm 0.03 \, \unit{nm}$ for the Al layer, whose corresponding roughness map is shown in the Supporting Information. \\
	
	\begin{figure} [ht!]
		\begin{center}
			\includegraphics[width=\textwidth]{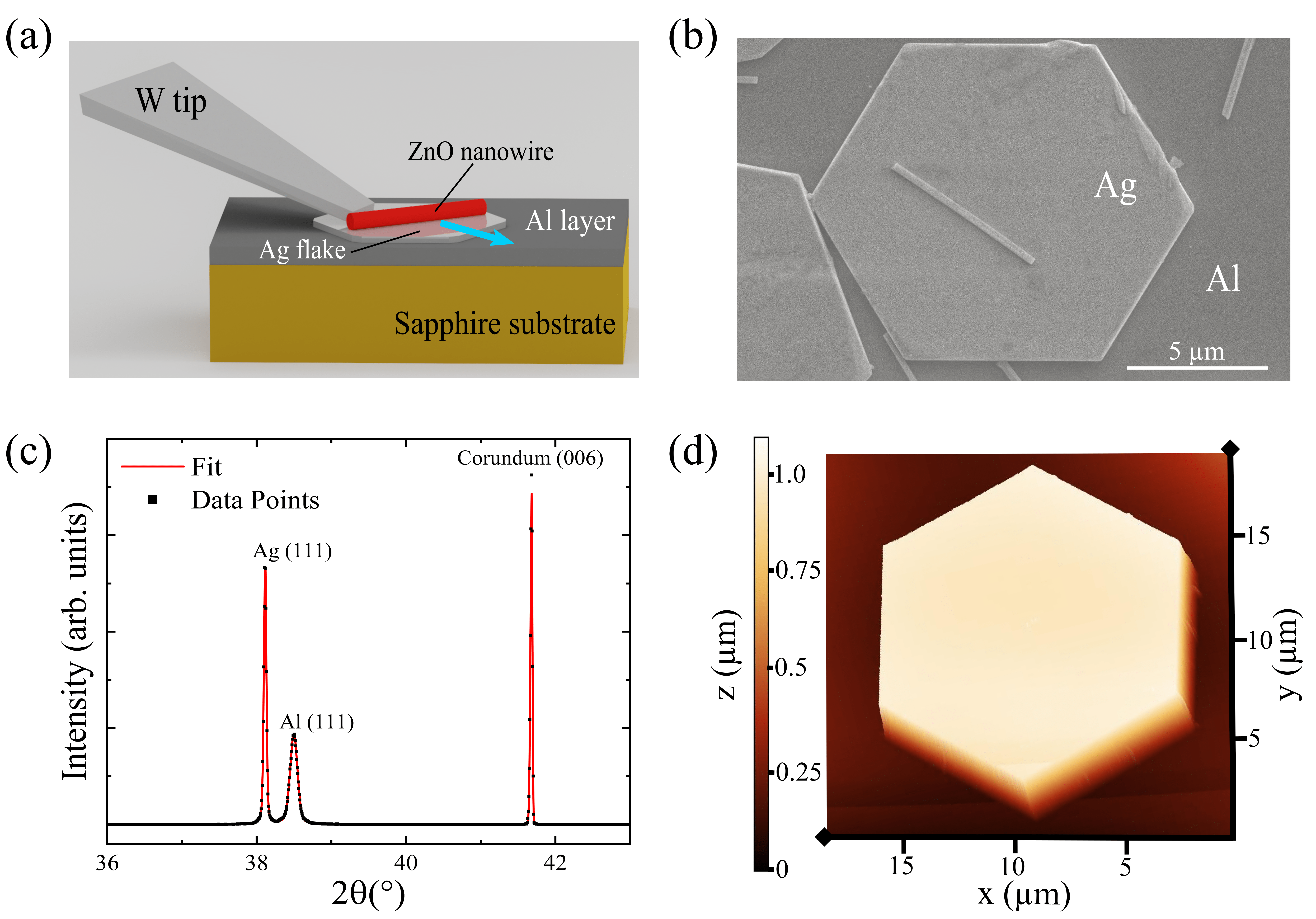}
			\caption{(a) Schematic of the W tip-aided manipulation of a nanowire moved off an Ag flake onto the Al layer during a single-nanowire $\upmu$-PL characterization. (b) SEM image of an exemplary ZnO nanowire imprinted on top of an Ag flake. (c) XRD spectrum displaying the (111) peaks for Ag and Al and the (006) peak of the sapphire substrate, fitted by pseudo-Voigt functions. The narrow full width at half maximum (FWHM) of the (111) peaks is indicative of a very high degree of crystallinity for both metals. (d) 3D AFM height map of a representative flake of lateral size $l$ $\approx$ 10 $\upmu$m, and thickness $t$ $\approx$ 1 $\upmu$m, evidencing a low surface roughness, attested to be $R_q < 1 \, \unit{nm}$ .}
			\label{figure:fig1}
		\end{center}
	\end{figure}

	\textbf{Figure 2}(a) shows the room-temperature PL spectra as a function of the excitation power fluence (normalized to the threshold value) for a nanowire of $L \approx 3.6 \, \unit{\upmu m}$ and average diameter $\langle d \rangle \approx 160 \, \unit{nm}$, first lying atop the Ag flake (top panel), and subsequently moved onto the Al substrate (bottom panel). On both substrates, the nanowire PL spectra are characterized by the appearance of lasing modes above the threshold, with a more pronounced emergence of the emission peaks on the higher-energy side of the broad spontaneous emission spectrum, as a result of the large ohmic losses and the consequent band filling of the gain medium conduction band induced by the presence of excess carriers \cite{liu2013}. The lasing curves of Figure 2(b) evidence a clear decrease in the estimated threshold fluence for the nanowire on the Al substrate - from $F_{th} = 1.34 \pm 0.08 \, \unit{mJ \, cm^{-2}}$ on Ag to $F_{th} = 1.02 \pm 0.08 \, \unit{mJ \, cm^{-2}}$ on Al -, as an indication of the higher gain overlap with the cavity and the lower propagation losses for the modes sustaining the lasing action, compared to the Ag case. Likewise, the higher threshold for the ZnO/Ag system results from the larger field confinement at the metallic interface, typically resulting in larger Purcell factors compared to the case of a ZnO/Al system \cite{repp2023}. The one-to-one comparison of the lasing emission at an excitation density of twice the threshold for the two device configurations reveals a blueshift of $\Delta \lambda_0 \approx 1.4 \, \unit{nm}$ (12 $\unit{meV}$ for the central lasing wavelength $\lambda_0$ - extracted from the Gaussian fit of the background-subtracted lasing envelopes - for the nanowire on the Ag flake. Such blueshift can be attributed to band-filling processes and the concurrent increase of the gain medium chemical potential, due to the vicinity of the metallic layer \cite{roeder2016phd}. Concurrently, the necessity of pumping at higher fluences to compensate for the larger losses arising from the spectral vicinity of the Ag surface plasmon frequency leads to higher carrier concentrations in the gain medium conduction band. This results in larger gain provided at higher-energy states in the electron-hole plasma driving the population inversion, which leads to a more blueshifted lasing envelope \cite{liu2013,versteegh2011}.
	
	\begin{figure} [ht!]
		\begin{center}
			\includegraphics[width=\textwidth]{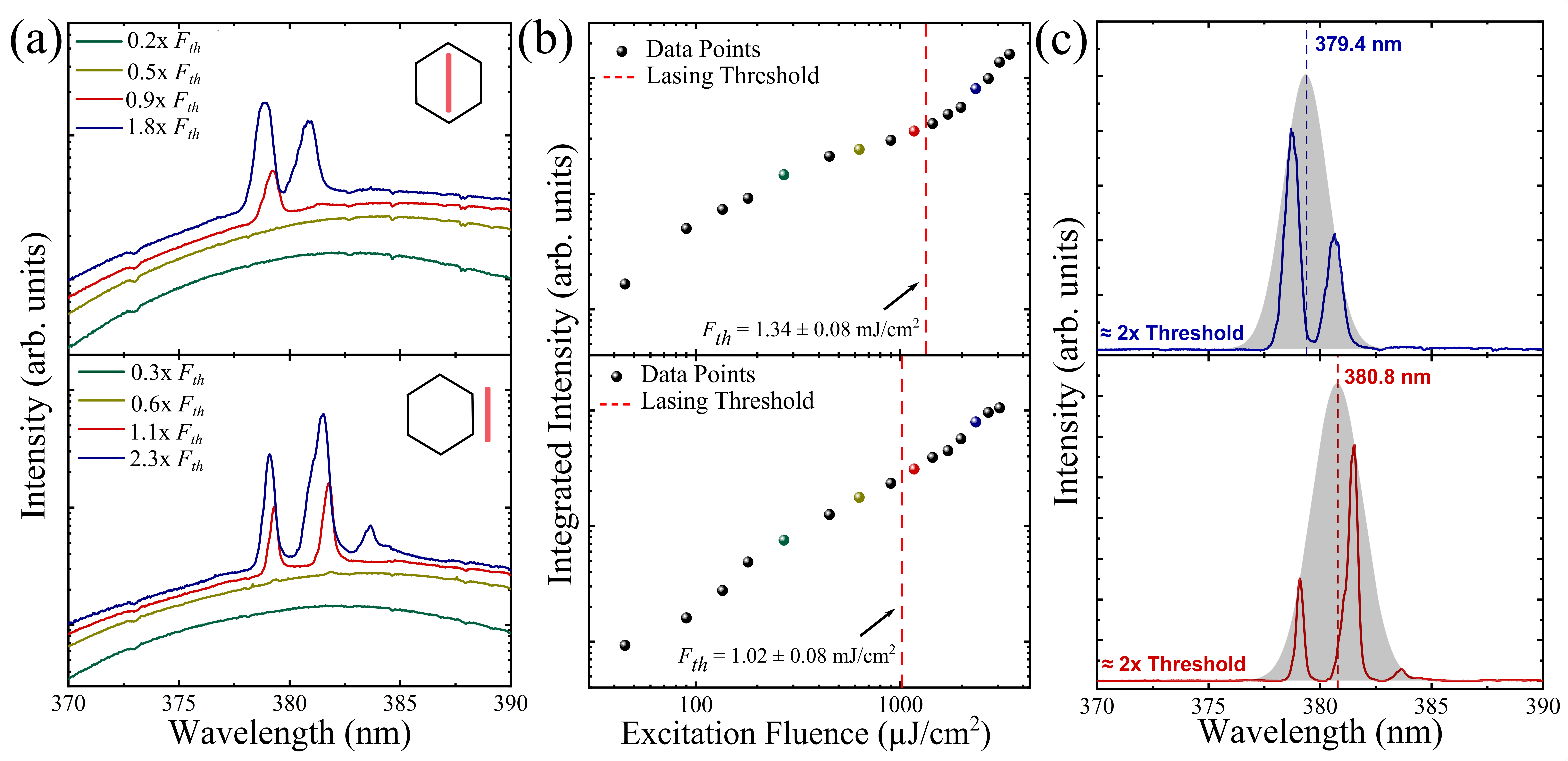}
			\caption{(a) Room-temperature log-scale PL spectra for the same nanowire of length $L$ $\approx$ 3.6 $\upmu$m and average diameter $\langle d \rangle$ $\approx$ 160 nm recorded on the Ag flake (top panel) and Al substrate (bottom panel). Both plots show evidence of lasing modes building on top of the spontaneous emission background with increasing pump fluence. (b) Corresponding double-logarithmic plots of the integrated spectral area of the stimulated emission as a function of the pump power on Ag (top panel) and Al (bottom panel). The dashed red lines mark the respective lasing thresholds, estimated by taking the fluence interval at which the integrated spectral areas of the stimulated emission and spontaneous emission are nearly equal. (c) Comparative PL spectra showing the Gaussian-fitted lasing envelopes.}
			\label{figure:fig2}
		\end{center}
	\end{figure}
	
	Additionally, the free spectral ranges $\Delta \lambda_{FSR}$ extracted by the PL spectra are $\Delta \lambda_{Ag} = 1.7 \pm 0.1 \, \unit{nm}$ and $\Delta \lambda_{Al} = 2.3 \pm 0.1 \, \unit{nm}$. Therefore, from the longitudinal mode spacing relation for an optical resonator:
	
	\begin{equation}
		\Delta \lambda_{FSR} = \dfrac{\lambda_0^2}{2n_gL}
	\end{equation}
	
	we could calculate the respective group indexes and found $n_{g,Ag} = 11.8 \pm 0.3$ and $n_{g,Al} = 8.8 \pm 0.2$, respectively. We also calculated the theoretical group index for a purely photonic cavity of the same length to be $n_{g,phot.} = 7.9 \pm 0.2$, based on the theoretical mode spacing $\Delta \lambda_{phot.} = 2.5 \pm 0.2 \, \unit{nm}$ obtained from the Fabry-Pérot dispersion relation:
	
	\begin{equation}
		\Delta \lambda_{FSR}^{theor.} = \dfrac{1}{L} \left[\dfrac{\lambda_0^2}{2}\left(n-\lambda_0^2\dfrac{dn}{d\lambda}\right)^{-1}\right]
	\end{equation}
	
	where $n \approx 2.2$ is the high-excitation renormalized refractive index of the gain medium \cite{versteegh2011} and $dn/d\lambda \approx -15 \, \unit{\upmu m^{-1}}$ the index dispersion at $\lambda \approx 380 \, \unit{nm}$ \cite{madelung2004,zimmler2010}. We can, thus, deduce that the group index found experimentally deviates from the theoretical value that a photonic device featuring the same resonator length would yield. Additionally, the higher value found for the nanowire on Ag is a result of the sudden change of the SPP dispersion around the excitonic resonance of ZnO, as a consequence of the sub-wavelength mode confinement \cite{chou2017}, while the group index for the ZnO/Al system approaches the theoretical value for a photonic device. The experimental findings also agree with the theoretical calculations for the group velocities found in reference \cite{repp2023}, showing a consistent increase for the ZnO/Al configuration compared to the ZnO/Ag one.

	\begin{figure} [ht!]
		\begin{center}
			\includegraphics[width=\textwidth]{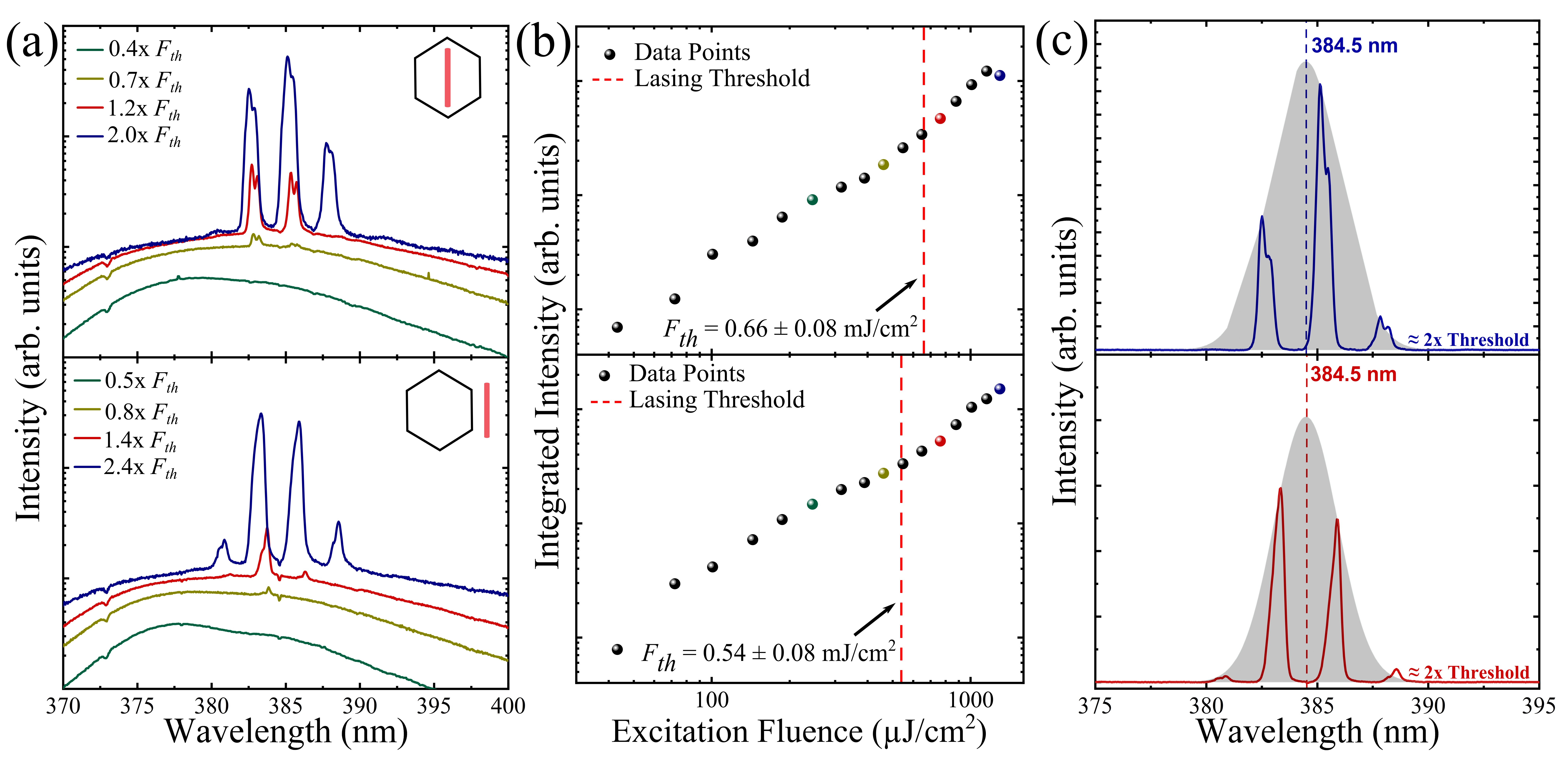}
			\caption{(a) Room-temperature log-scale PL spectra for a nanowire of length $L$ $\approx$ 3.8 $\upmu$m and average diameter $\langle d \rangle$ $\approx$ 190 nm recorded on the Ag flake (top panel) and the Al substrate (bottom panel). (b) Corresponding double-logarithmic plots of the integrated spectral area of the stimulated emission as a function of the pump fluence for the nanowire on the Ag flake (top panel) and on the Al substrate (bottom panel). The dashed red lines mark the respective lasing thresholds, estimated by taking the fluence interval at which the integrated spectral areas of the stimulated emission and spontaneous emission are nearly equal. (c) Comparative PL spectra showing the Gaussian-fitted lasing envelopes.}
			\label{figure:fig3}
		\end{center}
	\end{figure}
	
	To understand the diameter-constrained lasing characteristics, we investigated a second nanowire of comparable length ($L \approx 3.8 \, \unit{\upmu m}$) and larger diameter $\langle d \rangle \approx 190 \, \unit{nm}$, which is generally regarded as the theoretical limit below which a pure waveguide TE$_{01}$ mode loses its cavity confinement and its reflection coefficient drops abruptly \cite{zimmler2010}. Conversely, above this limit, such waveguide mode typically dominates the mode competition over HE$_{11}$ modes owing to the lower modal threshold gain \cite{sidiropoulos2014,roeder2016}. Moreover, the evanescent field of the lasing modes is usually negligible at larger diameters, due to their higher transverse confinement in the nanowire cavity \cite{maslov2003}, so we expect the influence of the different metallic substrates on the lasing of the thicker nanowire to be less significant compared to the thinner one. \textbf{Figure 3}(a) shows the corresponding room-temperature PL spectra as a function of the pump fluence (normalized to the threshold value) for the nanowire on the Ag flake (top panel) and Al substrate (bottom panel), respectively. The emergence of more Fabry-Pérot modes on the lower-energy side of the spontaneous emission background compared to the case of the thinner nanowire, typically regarded as a length-related resonator feature \cite{liu2013,zapf2017}, is, instead, indicative of a more likely diameter-driven photonic-like lasing behavior. As can be seen from Figure 3(b), this assumption seems to be confirmed by the overall decrease of the lasing threshold, compared to the thinner nanowire, which was, in turn, nearly invariant regardless of the metallic substrate. Indeed, the respective threshold fluences were estimated to be approximately 35\% and 50\% lower than those corresponding to the thinner nanowire coupled to the Ag flake and the Al substrate respectively, given a similar gain medium aspect ratio (namely  $ \langle d \rangle/L \approx \, 4\%$ for the thinner nanowire and $ \langle d \rangle/L \approx \, 5\%$ for the thicker one). The direct comparison of the Gaussian-fitted lasing envelopes at about twice the respective lasing thresholds, shown in Figure 3(c), reveals also a substantial invariance of the lasing central wavelength, irrespective of the metallic substrate. Like the case of the thinner nanowire, we calculated through \textbf{Equation (2)} the theoretical mode spacing for a photonic device featuring the same resonator length and found it to be $\Delta \lambda_{phot.} = 2.9 \pm 0.2 \, \unit{nm}$, thus, matching the experimental values extracted from the spectra $\Delta \lambda_{Ag} = \Delta \lambda_{Al} = 2.7 \pm 0.1 \, \unit{nm}$. Using \textbf{Equation (1)}, we could also estimate the group indexes and found that for a thicker nanowire, the group index of a ZnO/metal system approaches the theoretical value predicted for a corresponding photonic device, namely $n_{g,Ag} = n_{g,Al} = 7.2 \pm 0.2$ and $n_{g,phot.} = 6.7 \pm 0.2$. Such values are also in good agreement with those previously reported for nanowire cavities of comparable dimensions \cite{sidiropoulos2014}. 
	
	\begin{figure} [ht!]
		\begin{center}
			\includegraphics[width=\textwidth]{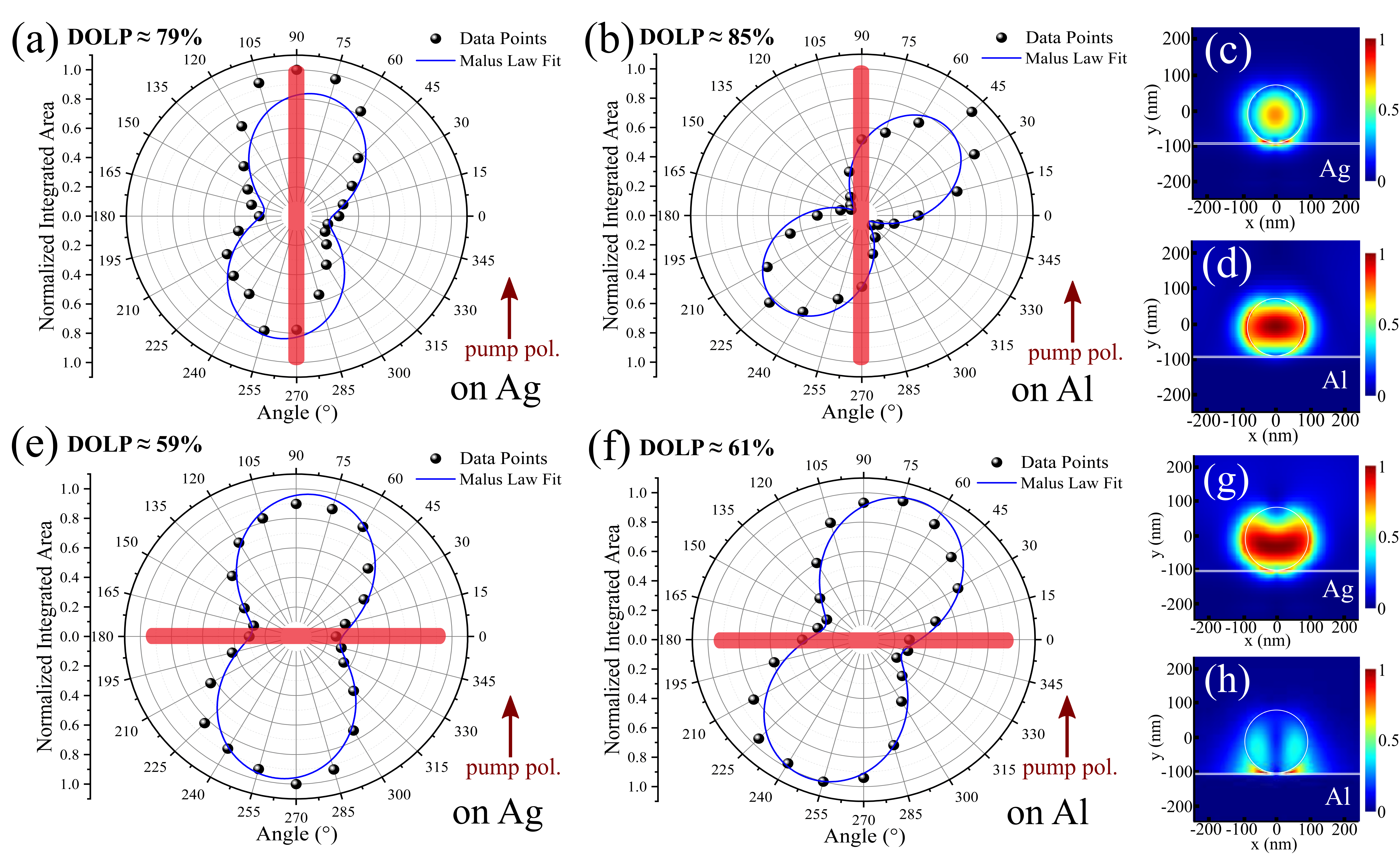}
			\caption{Polar plots for the lasing emission of the 160-nm thick ZnO nanowire on (a) the Ag flake and (b) the Al substrate, with the data points fitted by Malus’ law theoretical curve for retrieving the polarization direction relative to the nanowire orientation. The re-orientation of the emission polarization following the nanowire manipulation from the Ag flake onto the Al substrate, reveals a change in the mode competition processes sustaining the lasing action in the same nanowire as a function of the different metallic substrates. Simulated normalized modal field cross-sectional profiles for the 160-nm thick ZnO nanowire on the (c) Ag flake and (d) the Al substrate respectively, with the overlaid SIM cavity structure sketched in white. The polar plots for the lasing emission of the 190-nm thick ZnO nanowire on the (e) Ag flake and (f) the Al substrate, together with the corresponding normalized modal field cross-sectional profiles (g,h) are showed in the bottom panel.}
			\label{figure:fig4}
		\end{center}
	\end{figure}
	
	\textbf{Figure 4} displays the polar plots for the polarization-dependent measurements conducted at a pump fluence about twice the respective thresholds, for the (a) 160-nm thick nanowire excited on (a) the Ag flake and (b) the Al substrate. The polar plot in Figure 4(a) shows that for the nanowire on the Ag flake, the polarization of the stimulated emission is nearly parallel to the nanowire $c$-axis, as expected for a hybrid surface plasmon (HSP) mode characterized by a TM symmetry \cite{oulton2009,chou2018}. However, the corresponding FDTD-simulated modal field distributions shown in Figures 4(c,d) evidence also the appearance of an SPP-hybridized HE$_{11y}$ lasing mode within the nanowire cavity. This can be an indication of the fact that the actual mode sustaining the lasing action for a nanowire of $\langle d \rangle \approx 160 \, \unit{nm}$ on top of a silver substrate is rather characterized by a superposition of HSP and HE$_{11}$-like modes, both resulting in a lasing emission polarization parallel to the nanowire axis \cite{roeder2016}. On the other hand, there is an evident re-orientation of the polarization by $\approx$ 45° respective to the $c$-axis, after the same nanowire is moved onto the Al substrate. We attribute this to a lasing action sustained by the mode competition between HE$_{11x}$ and HE$_{11y}$ modes, which arises from the breaking of the mode degeneracy induced by the substrate \cite{roeder2016phd}. Such HE$_{11}$ modes are also partly coupling to the metal SPPs, as evidenced by the simulated modal field distributions in Figure 4(d). Therefore, the nanowire lasing emission polarization is the result of an intermediate state between SPP-hybridized HE$_{11}$ modes of different polarizations. The degree of linear polarization (DOLP) for both configurations was estimated from the relation:
	
	\begin{equation}
		DOLP = \dfrac{I_{max}-I_{min}}{I_{max}+I_{min}}.
	\end{equation}
	
	The polarization-dependent polar plots for the 190-nm thick nanowire, shown in Figure 4(e,f), evidence a stimulated emission predominantly polarized nearly orthogonal to the nanowire axis, regardless of the metallic substrate. The comparable orientation for the stimulated emission polarization suggests that, in both cases, the lasing action is dominated by the mode competition between HE$_{11x}$ and TE$_{01}$ modes, whose field distributions are shown in Figure 4(g,h). Namely, the HE$_{11x}$ mode dominates the lasing action for the on-Ag case, while a mixture of HE$_{11x}$ and TE$_{01}$ modes sustains the lasing for the on-Al case. In both cases, the field profiles evidence how the mode confinement at the metal interface becomes less and less efficient, in favor of a larger gain overlap of the lasing modes with the nanowire cavity: more details on the mode competition in the hybrid cavity can be found in the Supplementary Information. Additionally, we can notice that the lower values for the DOLP of the stimulated emission of the thicker nanowire compared to those found for the thinner nanowire might be ascribed to a less efficient pumping \cite{roeder2014,roeder2016phd} occurring for an orientation of the nanowire perpendicular to the pump polarization, compared to the case of a re-orientation of the nanowire parallel to the pump polarization (see Supplementary Information for further details).
	
	\subsection{Large-sample optical characterization of hybrid nanowire cavities}
	
	For the second part of the study, a high throughput $\upmu$-PL setup was used to investigate the lasing properties of a population of 7,434 ZnO nanowires \cite{church2023}. These nanowires were randomly distributed across the hybrid Ag flakes / Al substrate using dry imprinting: this resulted in some nanowires positioned atop the Ag flakes, and others placed onto the Al layer. A total of 6,933 nanowires on the Al layer were automatically located using bright-field optical imaging, and a further 501 nanowires atop the Ag flakes were manually identified.
	Whilst the above findings are stark for a single nanowire, it is important to establish whether the plasmonic enhancement effect of the Ag flakes is only significant for “champion” nanolasers, or if the effects persist across large nanolaser populations. This is a challenge due to observed variations in many nanowire properties (such as diameter, length, emission efficiency, plasmonic coupling strength, etc.) which is a consequence of the bottom-up growth and nanowire transfer technique. Therefore, to perform this assessment in a statistically robust manner, a large number of nanowires were studied, which enabled these effects to be averaged out in the analysis. This was achieved using a high-throughput $\upmu$-PL investigation: an example of each measurement on an individual nanowire is shown in the Supplementary Information, along with additional statistics, such as correlations among parameters. Firstly, the lasing yield was assessed, using power-dependent PL for a subset of 2,079 nanowires. As shown in \textbf{Figure 5}(a), 66\% of nanowires on the Al substrate underwent lasing, whereas only 54\% of the nanowires on the Ag flakes lased: these yields are similar to those found in previous studies using comparable excitation conditions \cite{sidiropoulos2014,chouBT2016}. The lasing yield depends upon the cavity dimensions \cite{repp2023,zimmler2010} and gain medium properties \cite{oulton2009,roeder2016phd}; however, since all the nanowires investigated in this study were sampled from the same population statistics, these properties follow the same distribution for nanowires on Al and Ag (see Supplementary Information). The lower lasing yield of the nanowires onto the Ag flakes is, thus, due to the impact of the flake itself, which can be ascribed to an increased plasmonic coupling resulting in higher lasing thresholds, as previously discussed. 
	
	\begin{figure} [ht!]
		\begin{center}
			\includegraphics[width=\textwidth]{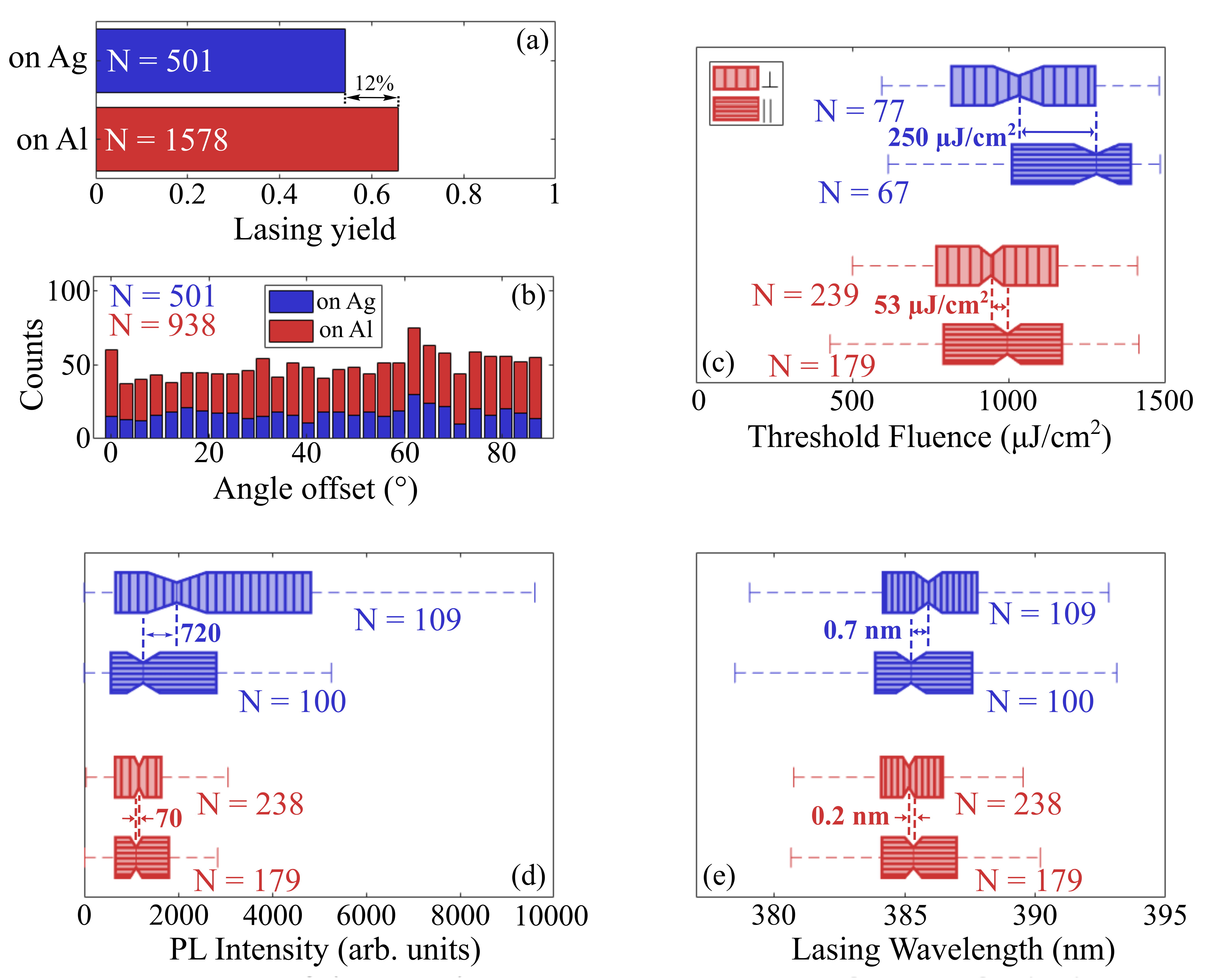}
			\caption{High-throughput $\upmu$-PL results, presented as histograms and boxplots, comparing nanowires on both Ag and Al substrates and with corresponding polarization directions (namely perpendicular and parallel to the nanowire c-axis). Histograms for (a) the nanowire lasing yields and (b) the angle offset between polarization and nanowires axis. Boxplots for the (c) threshold fluence, (d) low-excitation-power integrated PL intensity, and (e) peak wavelength of the dominant lasing longitudinal mode extracted from the PL spectra of the investigated nanolasers.}
			\label{figure:fig5}
		\end{center}
	\end{figure}
	
	A more detailed analysis was therefore explored, which utilized the polarization properties of the nanowire lasers. The single nanowire measurements described above demonstrate that the polarization direction of the lasing emission can be used as a metric for the plasmonic coupling strength: this direction was measured for a subset of the nanowires and is shown in Figure 5(b). This is a flat distribution with no preferred angle and no statistically significant difference between the two substrates. This result suggests that the nanowires provide a good sampling of plasmonic (0°), photonic (90°), and hybrid behavior given the variety of their diameters. Therefore, to investigate the differences between plasmonic and photonic lasing, the nanowires that are polarized parallel and perpendicular to the $c$-axis (within a 20° tolerance) were grouped into bins for further analysis.
	The impact of plasmonic coupling is demonstrated in Figure 5(c), which shows the lasing threshold fluence for these different groups. Nanowires on the Al substrate have median thresholds of 994(44) and 941(39) $\unit{mJ \, cm^{-2}}$ for parallel and perpendicular polarization, respectively, where the numbers in parentheses represent the 95\% confidence interval. Whilst the direction of this shift is compatible with the single nanowire measurements above (i.e. a more pronounced plasmonic coupling entails higher thresholds), the magnitude of the shift is not very significant - being at the 5\% level -, representing a median shift of 0.9$\upsigma$. On the Ag flakes, the median of the lasing threshold fluences has increased to 1279(73) and 1029(82) $\unit{mJ \, cm^{-2}}$ for the parallel and perpendicular polarizations, respectively. While the reason for the overall increase in threshold is not clear, the threshold shift between the two lasing characters (photonic- and plasmonic-like) is increased to 2.3$\upsigma$. This represents direct statistical evidence of the stronger plasmonic coupling between the nanowires and the Ag flakes.
	Further evidence of the enhanced plasmonic-like behavior of the nanowire cavities atop the Ag flakes can be obtained from a similar analysis of the PL intensity at low excitation powers. As shown in Figure 5(d), the median PL intensity of the nanowires on the Al substrate does not change for parallel and perpendicular polarization, respectively. These median intensities increase by 14\% and 80\% for the nanowires on Ag flakes using the same excitation. Since these nanowires are sampled from the same population, a change in the PL intensity distribution is plausibly related to a change in substrate. Indeed, moving from Al to Ag substrates, the increase in the median PL intensity could be ascribed both to the different surface reflectivity and plasmonic coupling between the two substrates. However, the impact of plasmonic coupling is clear when comparing the two polarization directions. For both substrates, the parallel polarization has a lower PL intensity, which we attribute to enhanced non-radiative recombinations resulting from increased polariton-phonon scattering processes. Indeed, the vicinity of the metal layer can produce intensified lattice vibrations in the gain medium, as a result of the strong electromagnetic field induced by the surface plasmons \cite{zhao2020}. This effect is relatively small on Al, with a significance of only 0.4$\upsigma$ at the 5\% level – but the results suggest that plasmonic coupling is stronger for the Ag flakes, with a significance of 1$\upsigma$. However, the interquartile range on the Ag flakes is about 2-3 times higher than on the Al layer, which may reflect a larger degree of variation in the enhanced plasmonic coupling efficiency, also as a function of the nanowire diameter \cite{repp2023}. 
	Finally, the individual nanowire measurements demonstrate that an enhanced plasmonic coupling can result in a blueshift of the lasing emission, due to band-filling effects \cite{liu2013,roeder2016phd}. Figure 5(e) shows the statistics for the wavelength of the dominant longitudinal mode for each substrate and polarization configuration. This demonstrates that there is no statistically significant shift in the lasing peaks on the Al layer when comparing photonic- and plasmonic-like characters. Conversely, but in line with the measurements of the individual nanowires, there is a small blueshift of 0.7 nm (with a statistical significance of 0.8$\upsigma$) of the plasmonic Ag-flake/nanowire lasers when compared to the more photonic-like counterparts.
	
	\section{Conclusion}
	
	The single-nanowire PL measurements demonstrated that the Ag flakes provide an enhanced plasmonic character, compared to the Al substrate, to the lasing of a nanowire with a diameter of 160 nm, namely, close to the theoretical cutoff for waveguide modes. This is evidenced by an increased lasing threshold, due to the high field confinement at the metal interface, a blueshifted lasing emission, due to band-filling processes, and a linear polarization that is parallel to the nanowire axis, as a result of a lasing action sustained by HSP and SPP-hybridized HE$_{11}$ modes. However, these effects are already mitigated for a nanowire with a 30-nm larger diameter, as a consequence of the less efficient coupling between the exciton-polaritons and the metal surface plasmons. 
	To address the substrate influence on a statistically more reliable basis, a large sample of 7,434 nanowires from the same growth batch was chosen for the high-throughput PL measurements, of which more than 2,000 were successfully driven to lasing. The results confirmed that the nanowires on Ag flakes exhibiting lasing action, consistently featured an enhanced plasmonic coupling between the nanowire cavity and the metallic substrate, evidenced by a polarized emission along the nanowire axis, an increase in the median lasing threshold, a reduction in the median low power PL intensity and a blueshift in the median lasing wavelength. 
	These findings suggest that the choice of the metallic substrate and the nanowire dimensions are crucial in determining the performance of UV nanolasers. Both substrate dispersive behavior and nanowire diameter are, indeed, interconnected through the degree of coupling between excitons and surface plasmons. This coupling defines, in turn, the extent to which more plasmonic- or photonic-like characteristics can influence the lasing properties of such nano-sources of coherent radiation. Therefore, these guidelines are to be carefully taken into account when designing nanowire-based SIM laser structures, in view of their envisaged integration into all-optical circuitry and short-wavelength hybrid waveguides.

	\section{Experimental Section}
	
	\subparagraph*{Synthesis of ZnO nanowires}
	
	The ZnO nanowires used in this work were grown via a vapor-liquid-solid (VLS) method inside a low-pressure horizontal three-zone tube furnace heated up to $1050\unit{\celsius}$ \cite{borchers2006}, by making use of a ZnO/graphite (8:1 molar ratio) source powder and Si substrates coated with an Au catalyst layer.
	
	\subparagraph*{Synthesis of the Ag flakes onto the Al substrates}
	
	The aluminum layer was grown via molecular beam epitaxy (MBE) onto a sapphire substrate \cite{liu2015,cheng2018}, featuring a thickness of $t = 96 \pm 1 \, \unit{nm}$. Further details on the growth can be found in the Supporting Information . The Al film was subsequently used as a substrate for the in-situ synthesis of Ag flakes via a metol-based reduction of silver nitrate (AgNO$_3$) according to the protocol reported in reference \cite{schoerner2021}. The Ag flakes exhibited a mean lateral size of $l = 8.9 \pm 5.4 \, \unit{\upmu m}$, with some flakes being also as large as $l \approx 30 \, \unit{\upmu m}$. Their mean thickness was $l = 693 \pm 404 \, \unit{nm}$, with a few flakes as thin as $l \approx 100 \, \unit{nm}$. The Al substrate and the Ag flakes were coated with a 3-nm thick Al$_2$O$_3$ spacer via atomic layer deposition (ALD) (see Supplementary Information for further details). 
	
	\subparagraph{Optical characterization}
	
	The room-temperature optical characterization of individual nanowires in the single-nanowire scheme was accomplished by defocusing the radiation emitted by a frequency-tripled Nd:YAG laser ($\lambda_{exc} = \, 355 \unit{nm}$ $f_{rep} = 100 \, \unit{Hz}$, $\tau_{pulse} = 7 \, \unit{ns}$) through a 100x NUV objective ($NA$ = 0.53) down to a spot size of 10 $\unit{\upmu m}$, so to illuminate the whole nanowire. The spectral acquisition was carried out by using a spectrometer (Princeton Instruments SP-2500i) equipped with a 1200 lines/mm grating (blazed at 300 $\unit{\nm}$), and connected to a nitrogen-cooled, front-illuminated CCD camera. The polarization-dependent PL emission was analyzed by placing in the detection path a NUV-optimized linear polarizer that was fully rotated by 360° in steps of 15°, via a motorized rotation mount. Both the power- and the polarization-dependent PL measurements were accomplished in the same configuration, namely with the pump polarization kept parallel to the nanowire $c$-axis, when not otherwise specified. For the large-sample optical characterization enabled by the high-throughput $\upmu$-PL spectroscopy, each nanowire was studied using several automated experiments, with excitation provided by a frequency-tripled ultrafast Yb:YAG laser ($\lambda_{exc} = \, 343 \unit{nm} \, f_{rep} = 200 \, \unit{kHz}, \, \tau_{pulse} = 150 \, \unit{fs}$). These experiments included optical bright field imaging using a 20x ($NA$ = 0.22) objective and low-power (sub-threshold) PL with a spot size of 4 $\unit{\upmu m}$ and an average power of 1 $\unit{\upmu W}$.  The lasing properties were assessed on a subset of 2,075 nanowires using power-dependent PL with a 50x NUV objective ($NA$ = 0.5) and a laser spot defocused to a diameter of 16 $\unit{\upmu m}$. Emission imaging below and above the threshold was also performed for each nanowire using the same excitation configuration. The polarization-dependent emission was assessed using an automated linear polarizer in the detection path. The spectral acquisition was achieved using a Horiba iHR550 spectrometer with a 1200 lines/mm grating blazed at 330 $\unit{nm}$ and an air-cooled front-illuminated CCD camera. For this high-throughput study, the pump polarization was not controlled and, thus, was randomly oriented relative to the nanowire $c$-axis.

	\medskip
	\textbf{Acknowledgements} \par 
	
	F.V., C.R., D.R., T.P., K.S.S., J.-S.H. acknowledge the financial support via projects C5 and C1 within the CRC 1375 “NOA-Nonlinear optics down to atomic scales” funded by the Deutsche Forschungsgemeinschaft (DFG) under Project ID: 398816777.  S.A.C. and P.P. acknowledge funding from a UKRI “Future Leaders Fellowship [MR/T021519/1]” and from the EPSRC (UK) [EP/V036343/1]. T.-H.D. and S.-D. L. acknowledge the National Science and Technology Council under grant No. 111-2221-E-A49-141-MY3. J.S.H. and K. S. S. thank the support from DAAD via PPP-Taiwan (Project ID: 57700890). 
	
\medskip
	
	\textbf{Conflict of Interest}
	
	The authors declare no conflict of interest.

	\newpage
	
	\section*{Supplementary Information}
	
	\section*{Sample Fabrication}
	
	\subsection*{Molecular Beam Epitaxy (MBE) Growth of Al Substrates}
	
	The epitaxial Al film was grown onto a sapphire substrate through a Varian Gen-II III-V solid-source MBE system,  employing an aluminum source with a 99.9998\% purity. The sapphire substrate was loaded into the MBE chamber and baked at $200\unit{\celsius}$ at the exit/entry chamber for 8 hours to remove moisture on the surface. Subsequently, the substrate underwent a 5-hour heating cycle at $500\unit{\celsius}$ in the preparation chamber to eliminate organic residues. Next, the sapphire substrate was transferred to the growth chamber, where it was heated to $650\unit{\celsius}$ for $1 \, \unit{h}$ and then gradually cooled below $0\unit{\celsius}$. The Al films were then grown on the treated sapphire substrate at a growth rate of $0.1 \, \unit{nm \, s^{-1}}$. The chamber maintained a vacuum level of the order of  $10^{-10} \, \unit{Torr}$ throughout the growth process. \emph{In-situ} characterization of both the substrate surface during pre-treatment and the Al film surface during growth was performed using reflection high-energy electron diffraction (RHEED). Further details regarding the growth process and structural characteristics of the Al films can be found in reference \cite{yeh2023}.

	\subsection*{Synthesis of Ag Flakes}

	\paragraph{Materials}
	
	Silver Nitrate (AgNO3, $\geq 99\%$, Merck) and metol (C$_7$H$_9$NO$\cdot$0.5H$_2$SO$_4$, $\geq 99\%$, Sigma-Aldrich).

	\paragraph{Synthesis}
	
	The synthesis of silver flakes followed a well-established protocol for single-crystalline silver flakes described by Schörner \emph{et al.} in reference \cite{schoerner2021}. The aqueous solutions of AgNO$_3$ and metol were prepared by dissolving $100 \, \unit{mg}$ of AgNO$_3$ and $100 \, \unit{mg}$ of metol individually in $3 \, \unit{ml}$ of milli-Q water. Since the solubility of metol in water is low, a 10-min sonication was employed to facilitate its dissolution. A beaker containing $20 \, \unit{ml}$ of milli-Q water with cleaned Al substrates placed vertically on the side wall was cooled in a refrigerator for $1 \, \unit{h}$ at $8\unit{\celsius}$. After $1 \, \unit{h}$, the beaker was taken out of the fridge, and $100 \, \unit{\upmu l}$ of freshly prepared AgNO$_3$ solution was added to it. Subsequently, $100 \, \unit{\upmu l}$ of metol solution was added drop-wise to the beaker within $10 \, \unit{s}$. After $60 \, \unit{s}$, the addition of $100 \, \unit{\upmu l}$ of the AgNO$_3$ and metol solutions was repeated three times with an interval of $60 \, \unit{s}$. Metol will reduce AgNO$_3$ to Ag and deposit on the substrate surface \cite{lyutov2014}. The appearance of visible silver flakes during the synthesis could be observed with the naked eye. Following the synthesis, the samples were carefully taken out of the beaker, immersed in an ethanol solution twice, and then dried under an N$_2$/air flow. The synthesized silver flakes were then coated, together with the Al layer, with a 3-nm thick Al$_2$O$_3$ spacer grown via atomic layer deposition, to protect them against oxidation.
	
	\subsection*{Atomic Layer Deposition (ALD) of the Oxide Spacer}
	
	The Al$_2$O$_3$  spacer was deposited by thermal atomic layer deposition (T-ALD) on the silver flakes and Al substrates using a commercial Oxford Instruments OpAL (Bristol, UK) ALD tool \cite{dathe2016,jang2020}. As precursors, Trimethylaluminum (TMA, EG-purity, Dockweiler Chemicals) and distilled water vapor were used. The thermal process was more sensitive than a plasma-enhanced ALD - using reactive oxygen instead of water vapor - and did not oxidize the underlying silver flakes. The dose times of TMA and water vapor were $12 \, \unit{ms}$ and $15 \, \unit{ms}$ respectively, followed by a purge time of $5 \, \unit{s}$ after each precursor dosing. The substrate temperature was set to $80\unit{\celsius}$ and a growth rate of $0.08 \, \unit{nm}$ per cycle was determined. For the deposition of a 3-nm thick Al$_2$O$_3$ spacer, 62 cycles of T-ALD were performed.
	
	\begin{figure} [ht!]
		\centering
		\includegraphics[scale=0.5]{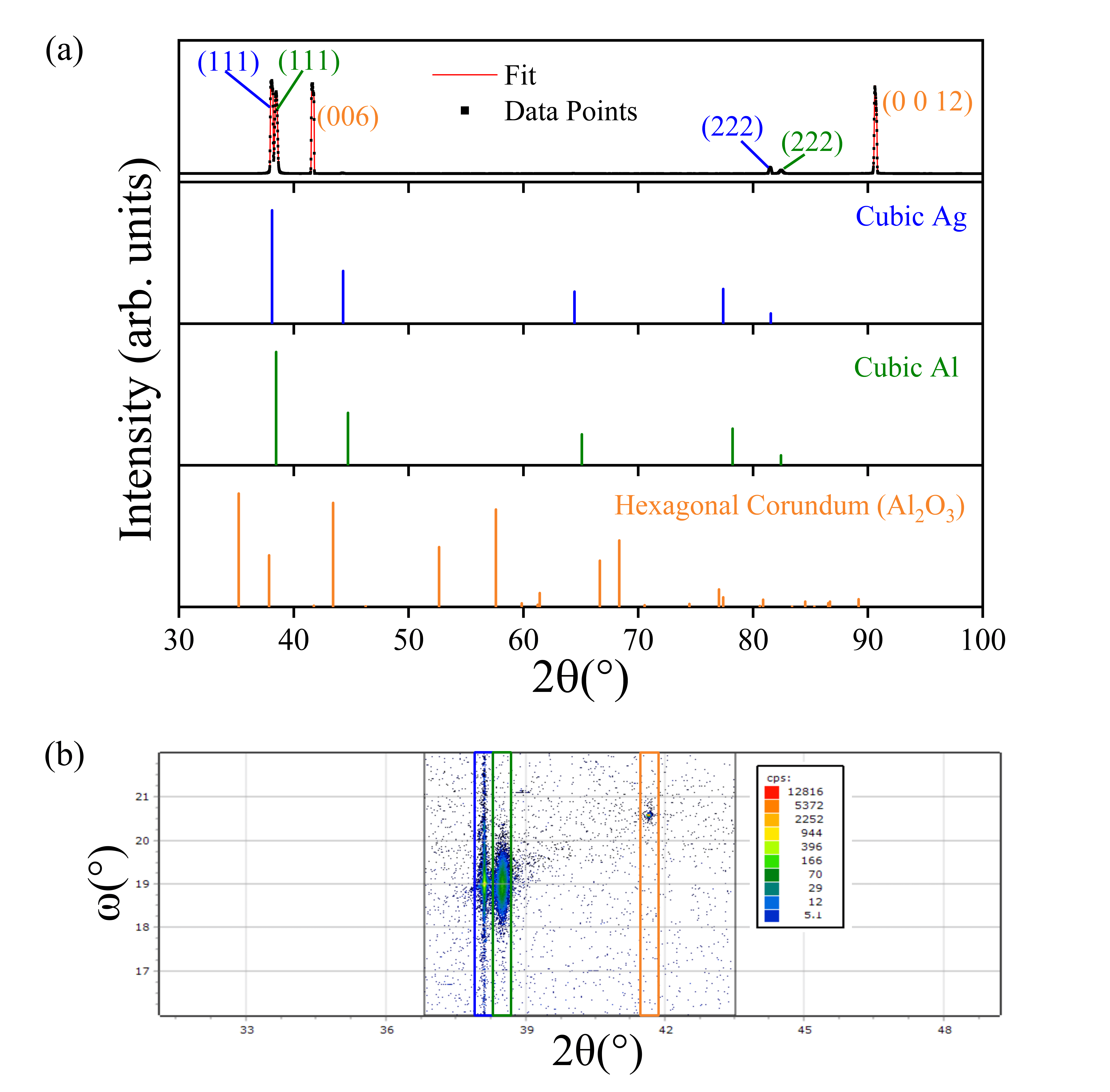}
		\caption{(a) Full-range XRD plot of a sample featuring Ag flakes on top of an Al layer, grown via MBE onto a sapphire substrate, with the corresponding indexed peaks. The sharp Al (111) peak evidences the high directionality of the growth and the remarkable crystal quality. (b) Corresponding RSM for the Ag (111) peak (blue), Al (111 peak (green) and Al$_2$O$_3$ (006) peak (orange), with respective intensities plotted on a logarithmic scale.}
		\label{figure:XRD}
	\end{figure}

	\section*{Sample Characterization}
	
	\subsection*{X-ray Diffraction (XRD) analysis}
	
	XRD measurements were accomplished with a Panalytical X'Pert Pro MPD theta-theta-diffractometer with a parallel beam setup for X-ray diffraction (XRD), grazing incidence XRD (giXRD) and X-ray reflectometry (XRR). The beam from a Cu long-fine-focus X-ray tube operated at $40 \, \unit{mA}$ and $40 \, \unit{kV}$ was directed through a parallel mirror with 0.5° divergence slit, 4-mm mask, 0.04-rad Soller slit, and 1.4-mm anti-scatter slit aligned to the horizontally-positioned sample. The height of the sample was adjusted to the goniometer-circle. The XRD measurements were accomplished from 10° to 90° with 0.026° steps at a rate of $97 \, \unit{s}$ per step, while the giXRD measurements were carried out from 10° to 90° with 0.026° steps at a rate of $0.4 \, \unit{s}$ Phase analysis was carried out using the software MalvernPanalytical Highscore Plus v.4.9 \cite{degen2014} and comparing the XRD lines with reference cards from ICSD and COD databases.
	
	\begin{figure} [ht!]
		\centering
		\includegraphics[scale=0.35]{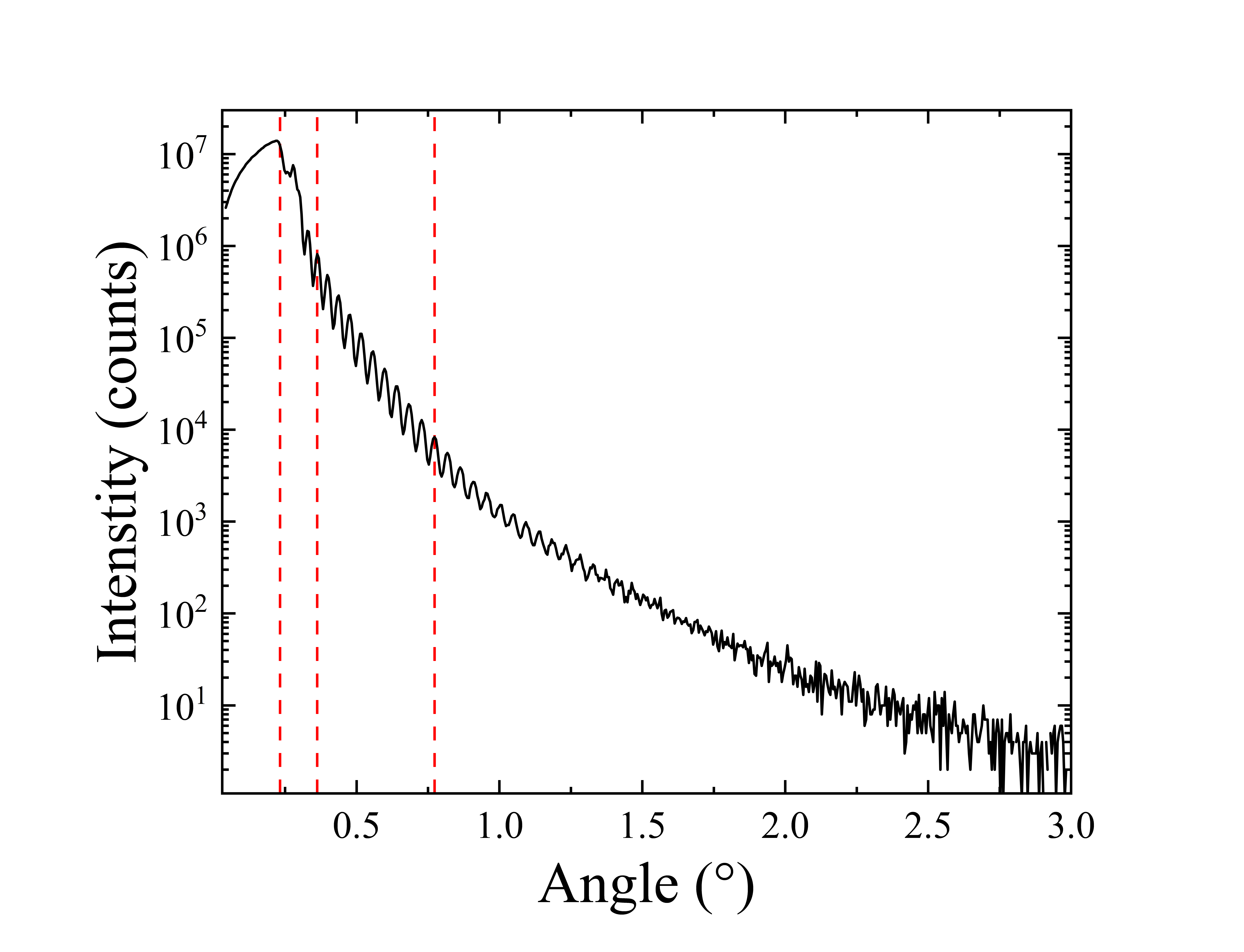}
		\caption{XRR spectrum of a bare Al film on a sapphire substrate. The dashed lines mark the corresponding critical angles used for the calculations of the film thickness.}
		\label{figure:XRR}
	\end{figure}

	To characterize the crystalline quality of the Ag flakes a hybrid monochromator (HM) was used in the incident beam with a 0.5° divergence slit and a 4-mm mask. Figure \ref{figure:XRD}(a) shows 3 narrow peaks - from left Ag(111), Al(111) and Sapphire (006) - around 40°, as presented in the main text of the manuscript and fitted by pseudo-Voigt functions, as well as the (222) peaks around 80° and the Sapphire (00 12) peak around 90°. With the HM and the PIXCEL3D detector, a reciprocal space map (RSM), shown in Figure \ref{figure:XRD}(b) was recorded. The scan axis was set to $\omega$ from 16° to 22° with 0.0065° steps at a rate of $97 \, \unit{s}$ per step; the plot was exported from MalvernPanalytical AMASS v.1.1.

	\subsection*{XRR Analysis}
	
	An X-ray reflectance (XRR) scan was also carried out from 0.04° to 3° with 0.005° steps at a rate of $3 \, \unit{s}$ per step. Due to the high surface smoothness of the sapphire substrate and Al film, Kissing fringes could be observed. The fringe width was correlated with the film thickness, while the fringe depth with the surface smoothness. The reflectivity scan was analyzed with the software Malvern Panalytical AMASS v.1.1. The program allowed for the “manual” determination of the thickness by positioning three cursors at the critical angle (at which the X-ray beam starts to penetrate the sample) and at the first and second fringe angles. The fringe width was correlated with the thickness and the number of fringes used for the calculation, so a Fourier transform could be performed, yielding a peak at the film thickness. From the XRR plot and its corresponding Fourier transform, one obtains a thickness of the MBE-grown Al layer of $t = 96 \pm 1 \, \unit{nm}$.

	\section*{Atomic Force Microscopy (AFM) Anaylsis}
	
	\begin{figure} [ht!]
		\centering
		\begin{subfigure}{.75\textwidth}
			\includegraphics[width=\linewidth]{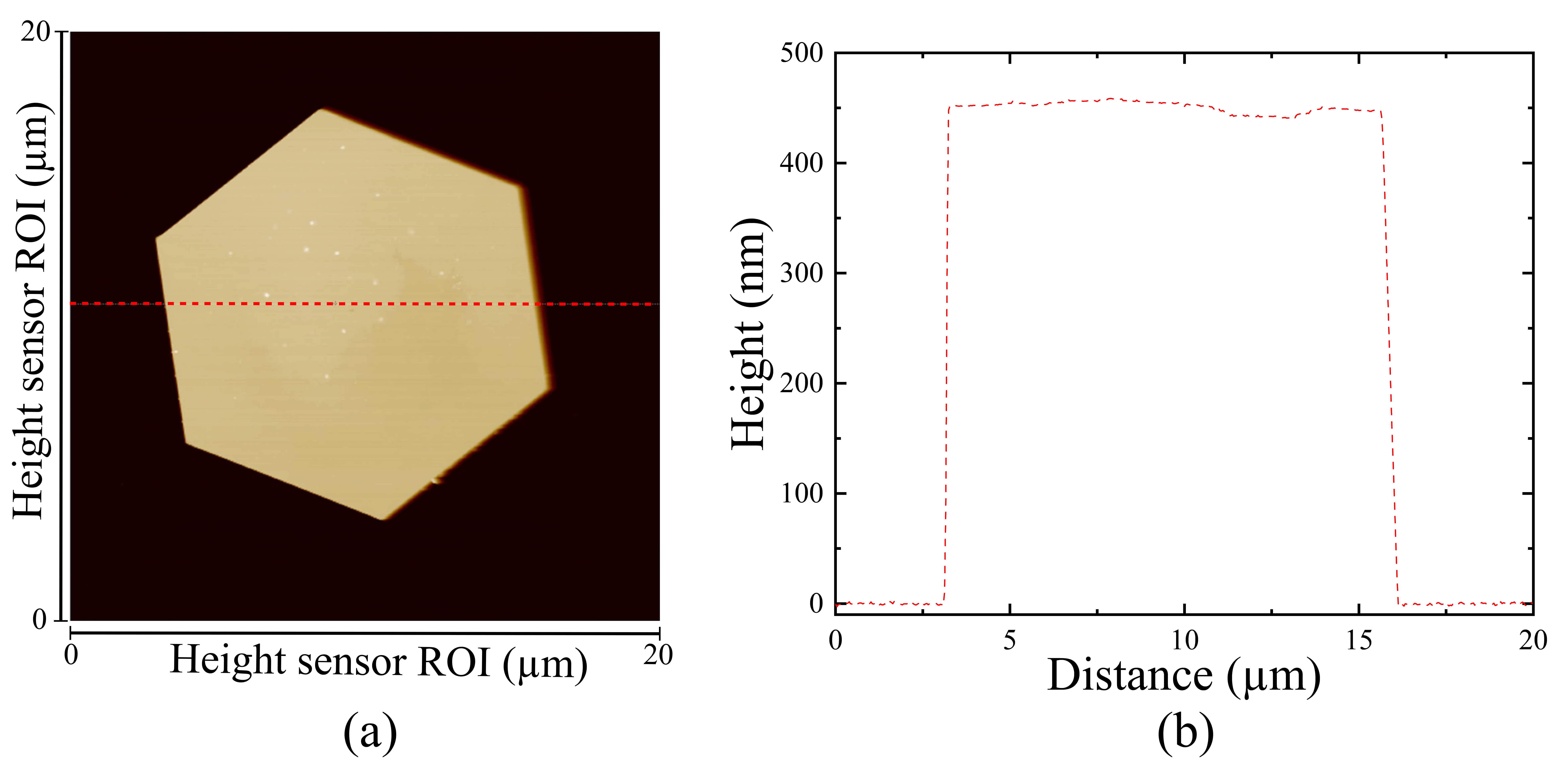}
			\phantomsubcaption
		\end{subfigure}
		\\
		\begin{subfigure}{.75\textwidth}
			\includegraphics[width=\linewidth]{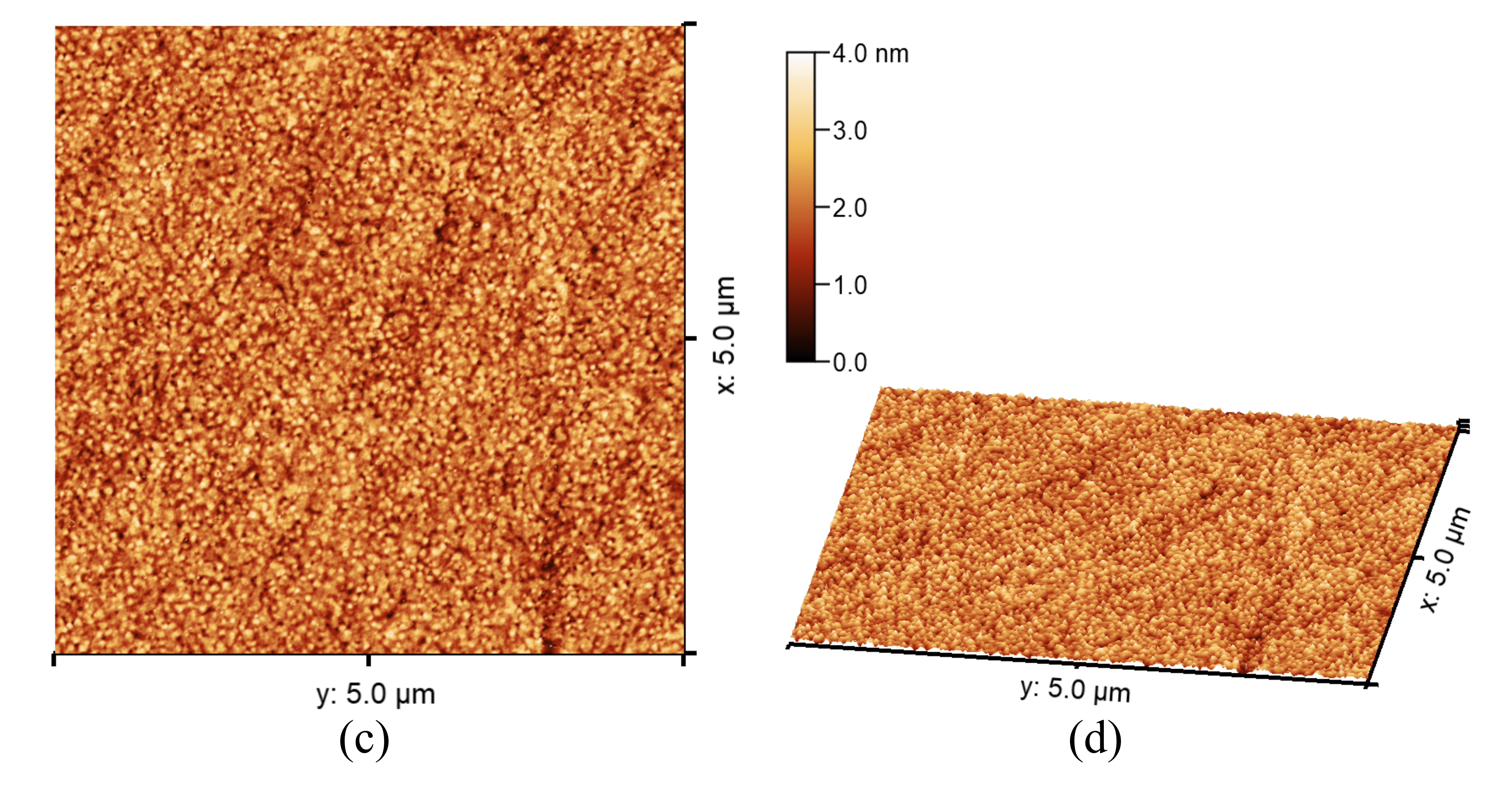}
			\phantomsubcaption
		\end{subfigure}
		\caption{AFM height map (a) and plot (b) for an exemplary Ag flake of lateral dimension $d \approx$ 7.5 $\upmu$m, measured within a 20x20 $\upmu$m ROI. The dashed red line marks the height sensor trace. AFM height 2D (c) and 3D (d) maps for the Al substrate measured within a 5x5 $\upmu$m ROI.}
		\label{figure:AFM}
	\end{figure}
	
	
	The AFM analysis was accomplished by a Bruker Dimension Edge system (BrukerNano surfaces) equipped with a 100-$\upmu $m scanner and enabling a $z$-range of up to $10 \, \unit{\upmu m}$. The measurements were performed in contact mode using a supersharp probe SNL (Bruker AFM probes),specified with a height of $5 \, \unit{\upmu m}$ and a tip radius of $2 \, \unit{nm}$. Depending on the different sizes of the flakes the scan range varied between 15, 20, and 40 $\upmu$m with a fixed $x,y$-resolution of 512x512 pixels. 
	
	The image evaluation was carried out using a 2D flattening procedure and map reconstruction by means of the open-source software Gwyddion v.2.65 \cite{necas2012}. For the determination of the surface roughness of the Ag flakes, as for the one representatively shown in Figure \ref{figure:AFM}(a,b), 6 scans within a 5x5 $\upmu$m region of interest (ROI) were measured, whereas for the roughness of the substrate, 5 scans were evaluated. After a 2D flattening procedure, the average root-mean-square roughness value was determined to be $R_q = 0.93 \pm 0.49 \, \unit{nm}$ for the measured Ag flakes and $R_q = 1.01 \pm 0.06 \, \unit{nm}$ for the Al substrate, whose exemplary height maps are shown in Figure \ref{figure:AFM}(c,d),

	\section*{Optical Characterization}
	
	\subsection*{Manipulation-Assisted Single-Nanowire PL Measurements}
	
	\begin{figure} [ht!]
		\centering
		\includegraphics[width=\textwidth]{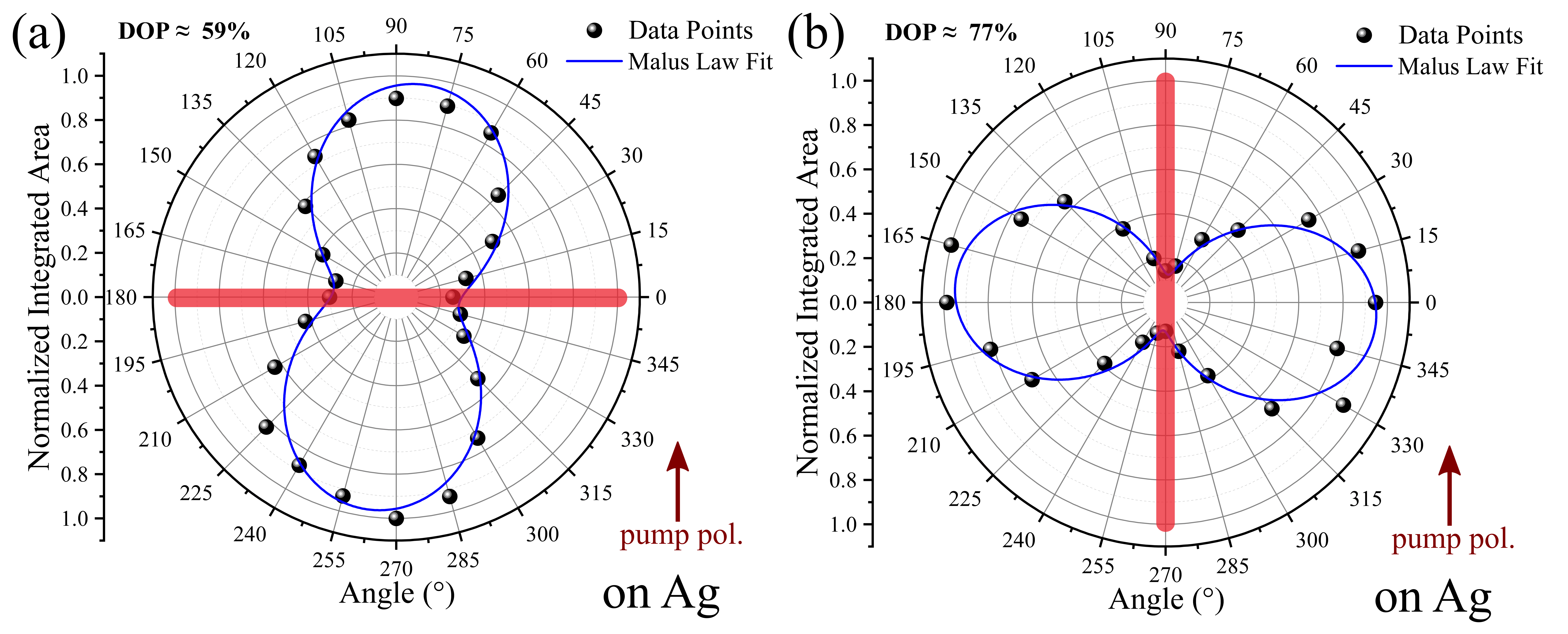}
		\caption{Polar plots of the on-flake stimulated emission polarization for the 190-nm thick nanowire fitted with a Malus’ law-based curve for an orientation of the nanowire axis (a) perpendicular and (b) parallel to the pump polarization. The different DOP can be attributed to the different pumping efficiency for the two orientations.}
		\label{figure:POLDEP}
	\end{figure}

	The polarization-dependent PL measurements within the manipulation-aided single-nanowire experiments were carried out by aligning the nanowire both parallel and perpendicular to the pump polarization, using a rotation mount. The difference in the nanowire orientation seemed to only affect the pumping efficiency - as previously found in references \cite{roeder2016,roeder2016phd} - which may impact the ratio between polarized and unpolarized light outputted by the cavity.
	When considering the 190-nm thick nanowire emission characteristics described in the main text, one can notice that, for an orientation of the nanowire parallel to the pump polarization, the DOLP was comparable to the value obtained for the thinner nanowire onto the Ag flake, which was also aligned parallel to the pump polarization. Indeed, as can be seen from Figure \ref{figure:POLDEP}, the overall polarization for the stimulated emission was still proven to be orthogonal to the nanowire axis, even after a re-orientation of the nanowire relative to the pump polarization. This result demonstrates that the underlying coupling between exciton-polaritons and surface plasmons is not significantly affected by the orientation of the pump polarization with respect to the nanowire $c$-axis. Moreover, these findings ensure that the absence of control on the relative orientation between pump polarization and nanowire orientation did not relevantly influence the results of the high-throughput polarization-dependent study performed on > 2,000 nanowires, as discussed in the main text.

	\subsection*{High-Throughput PL Measurements}
	
	\begin{figure} [ht!]
		\centering
		\includegraphics[width=0.75\textwidth]{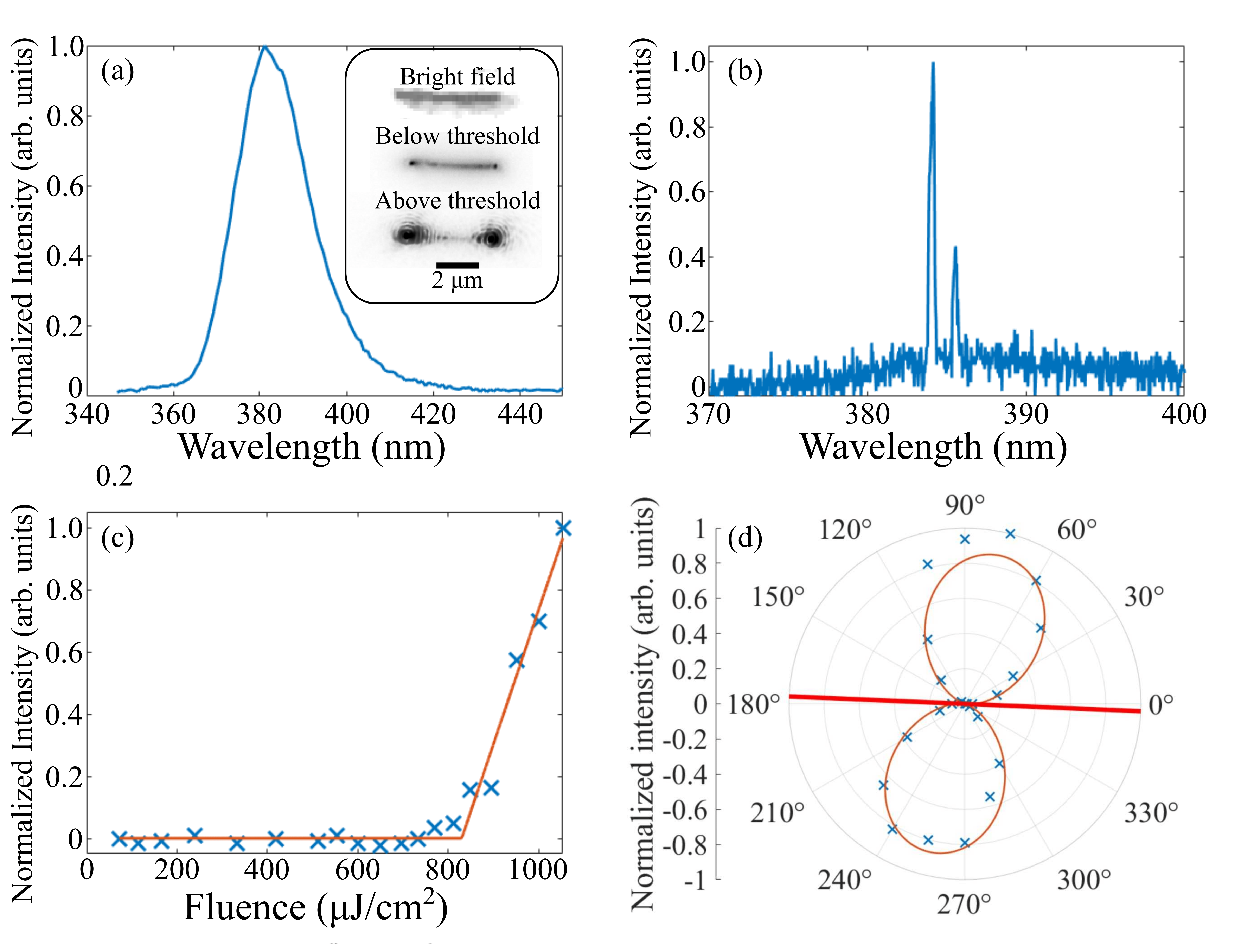}
		\caption{Results from the high-throughput $\upmu$-PL experiments from one single ZnO nanowire. (a) Low-power photoluminescence spectrum and imaging under bright field, below and above lasing threshold conditions (inset). (b) Emission spectrum above the lasing threshold. (c) Light-in-light-out curve of the integrated intensity of the dominant lasing peak. The data is fit with a threshold model with a linear fluence dependence above the threshold. (d) Polarization plot for the dominant lasing mode, fitted with Malus’ law to extract the DOLP and polarization angle.}
		\label{figure:Stat1}
	\end{figure}

	High-throughput $\upmu$-PL spectroscopy was used to perform 6 independent experiments on a subset of 7,434 nanowires that were randomly scattered across the Al substrate and Ag flakes. An example of the results on a single nanowire, chosen randomly from the full population, is shown in Figure \ref{figure:Stat1}. This includes low-power PL, displayed in Figure \ref{figure:Stat1}(a), which is typical of most studied nanowires and provides information regarding the luminescence intensity and bandgap \cite{church2023}. Fluence-dependent photoluminescence was used to assess the lasing performance. Sharp lasing peaks were observed in the spectrum when excited above the threshold fluence, as shown in Figure \ref{figure:Stat1}(b), and the intensity of the dominant peak, with a wavelength of 384 $\unit{nm}$ in this example, was used to assess the lasing threshold, as shown in Figure \ref{figure:Stat1}(c), giving a value of $F_{th} = 0.83 \pm 0.05 \, \unit{mJ \, cm^{-2}}$. More information on this analysis can be found in reference  \cite{church2022}. Finally, the nanowires which lased were investigated using polarization-resolved spectroscopy by placing a linear polarizer on the detection path and rotating this whilst recording spectra. As with the threshold, the intensity of the dominant lasing mode was used to plot intensity versus angle, as in Figure Figure \ref{figure:Stat1}(d), and to determine the degree of linear polarization (DOLP = 86\%). The orientation of the nanowire was determined from the imaging results, and, hence, the relative angle of the emission polarization was found to be $\approx$ 80° for this nanowire. As discussed in the main paper, it is therefore likely that the lasing of this nanowire has a predominant photonic-like character.
	Bright-field imaging allows a measurement of the nanowire length ($L \approx 5.8 \, \unit{\upmu m}$ in this case), but does not have the resolution to measure the diameter. When exciting below the lasing threshold, the majority of the spontaneous emission is detected uniformly along the nanowire, matching the homogeneous illumination, with a slight enhancement at the end facets due to waveguiding effects \cite{li2013}. Above the threshold, the vast majority of the emission is detected at the end facets due to strong coupling between the lasing mode and the lasing cavity \cite{zimmler2010}.

	\subsection*{Additional High-Throughput Statistics}
	
	\begin{figure} [ht!]
		\centering
		\includegraphics[width=\textwidth]{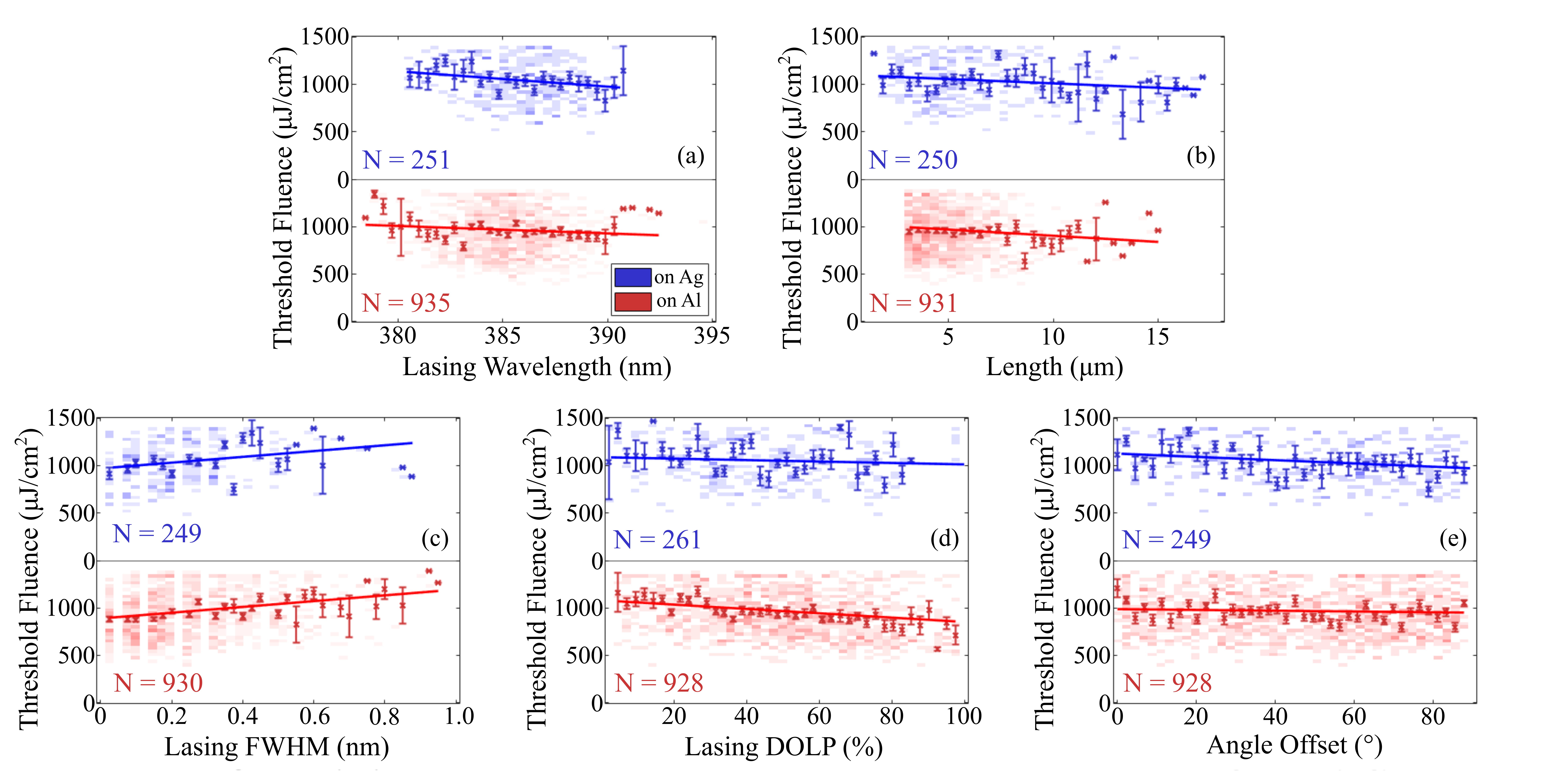}
		\caption{Two-dimensional histograms showing correlations between parameters and the lasing threshold fluence, that can be observed from the results of the high-throughput $\upmu$-PL experiments on Al (red) and Ag (blue) substrates. The threshold fluence is correlated with (a) lasing wavelength, (b) nanowire length, (c) lasing spectral width, (d) DOLP for the lasing emission and (e)  offset angle between lasing polarization and nanowire $c$-axis. The points and error bars represent the median values and standard error in every horizontal bin.}
		\label{figure:Stat2}
	\end{figure}
	
	
	The high-throughput experimental approach is advantageous when used to study materials where several properties are characterized by large variability and inhomogeneity \cite{church2023}. This is because, within the framework of small-scale studies, variations in these properties can obscure important trends and correlations. In this scenario, measuring thousands of individual devices using automation can provide a means to average over the multi-dimensional variation and extract the important relationships with statistical rigor.
	Figure \ref{figure:Stat2} demonstrates the effectiveness of this approach on a large-sample population of ZnO nanowires by correlating the factors that influence the lasing threshold on both substrates. These relationships are characterized by Pearson’s linear correlation coefficient $r$ and are assessed at the 5\% significance level. Figure \ref{figure:Stat1}(a) shows that lower thresholds occur for longer-wavelength nanolasers, with $r$-values of -0.07 and -0.16 for Al and Ag, respectively. This behavior for nanowire lasers has been typically attributed to a more photonic-like character of the lasing action sustained by waveguide modes, leading to a redshift of the emission via photon re-absorption/re-emission cycles induced by the presence of available states in the Urbach tails of the gain medium conduction band \cite{li2013,roeder2016phd}.  The fact that the correlation is stronger for the Ag flakes may reflect the stronger plasmonic coupling for a subset of nanowires of smaller diameter, characterized by a more pronounced blueshift, as discussed in the main text. 
	
	Figure \ref{figure:Stat2}(b) shows that the nanowire length is also critical, with $r$-values of -0.10 and -0.14 for Al and Ag, respectively. However, this can be well understood: a longer nanowire reduces the lasing threshold since more gain material is available for promoting the lasing transition. Figure \ref{figure:Stat2}(c) demonstrates a strong positive relationship between the full width at half maximum of the lasing peaks and the threshold, with $r$ values of 0.22 and 0.20 for Al and Ag, respectively. The width of the lasing peaks is related to several factors, such as material homogeneity \cite{zapf2019}, cavity quality factor \cite{couteau2015} and, due to the ultrafast excitation conditions, spectral shifting of the emission on recombination timescales \cite{sidiropoulos2014}, all of which are known to impact the lasing performance of the nano-devices \cite{church2022}.  Figure \ref{figure:Stat2}(d) shows the relationship between the degree of linear polarization of the lasing emission and the lasing threshold. The data trend suggests a strong relationship with an $r$-value of -0.21 for the Al substrates, but no statistically significant correlation for the Ag flakes. Variation in the DOLP can be attributed to the depolarizing effect of light scattering on the metallic substrate. These results suggest that, for the Al substrates, a reduction in this depolarizing effect, perhaps due to a lower average surface roughness compared to the Ag flakes, can lead to a decrease in the threshold, associated with reduced scattering and absorption losses. Finally, Figure \ref{figure:Stat2}(e) addresses the impact of the polarization direction, relative to the nanowire $c$-axis, on the lasing threshold, as an indication of the different lasing character. No correlation is observed for the Al substrate, but a strong negative correlation with an $r$-value of -0.19 is observed for the Ag flakes. This is an extension to the analysis in Figure 5 of the main text and is evidence for the enhanced plasmonic coupling between the ZnO nanowires and the Ag flakes compared to those on the Al substrate. This is demonstrated by the fact that, as the offset angle increases to 90°, the lasing character becomes more photonic-like, leading to a decrease in the threshold fluence.
	
	\section*{Finite-Domain-Time-Difference (FDTD) Simulations}
	
	\begin{figure} [ht!]
		\centering
		\includegraphics[width=\textwidth]{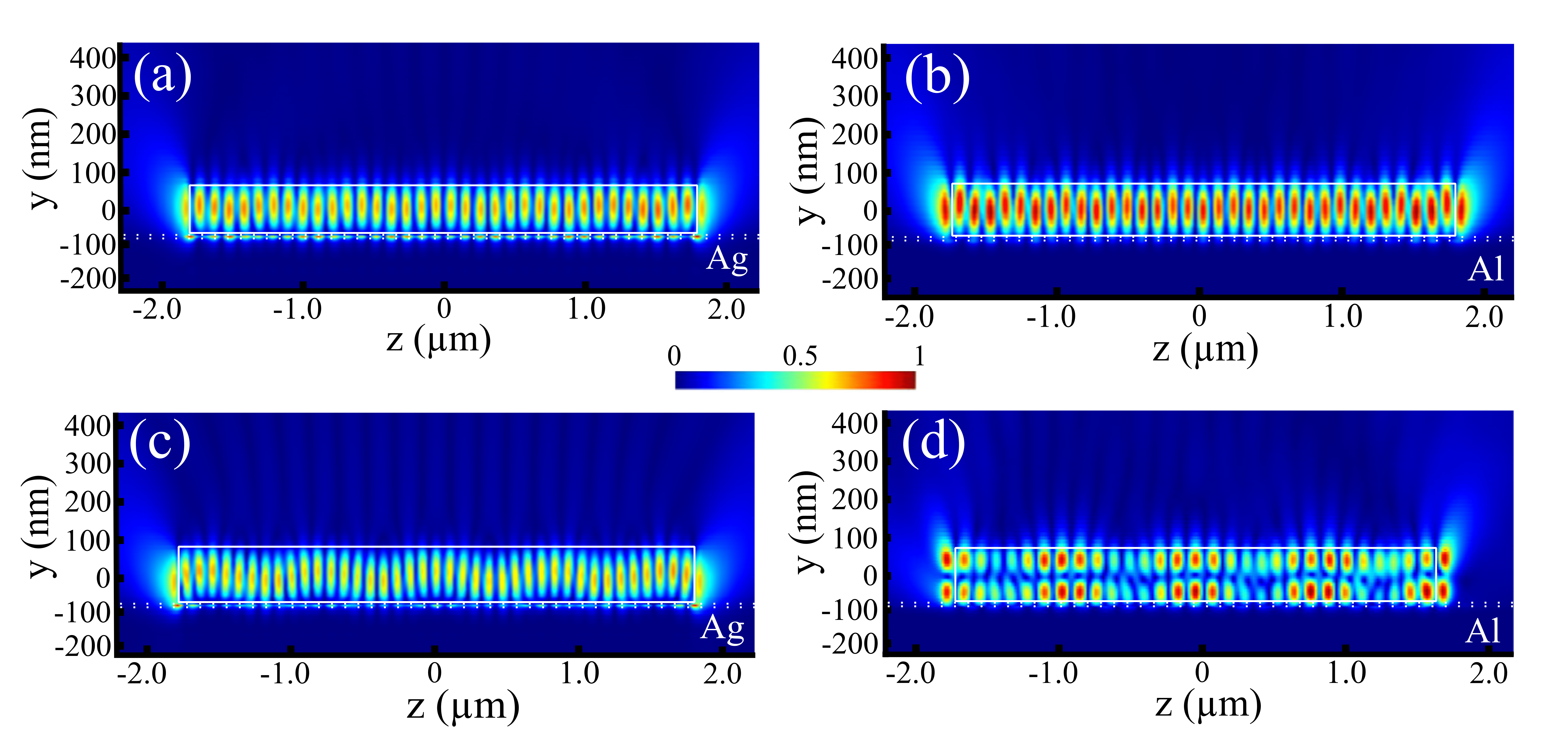}
		\caption{Longitudinal modal field distributions for the 160-nm thick nanowire (a) on Ag and (b) on Al. The color bar indicates the normalized values for the electric field amplitudes. and on the Al substrate for the (b) 160-nm thick and (d) 190-nm thick nanowires.}
		\label{figure:NWsimul}
	\end{figure}

	The field profiles were calculated via the finite-difference time-domain (FDTD) solver implemented in the commercial software suite Lumerical \cite{lumerical2021}. The nanowire material permittivity was derived from a model based on quantum field theory \cite{versteegh2011}, whereas the material parameters for the metallic substrates were taken from reference \cite{palik1998}, and the spacer was modeled as a dispersionless dielectric with a permittivity of $\varepsilon = 2$. For the lasing transition, Lumerical’s built-in 4-level-2-electron plugin material was used, where the lasing transition has a wavelength of $\lambda_{las} = 380 \, \unit{nm}$ and a width of $\Delta\lambda_{las} = 100 \, \unit{nm}$ and the pumping level has a wavelength of $\lambda_{exc} = 355 \, \unit{nm}$ and a width of $\Delta\lambda_{exc} = 10 \, \unit{nm}$. The inversion of the lasing transition was set to 0.1, which is in good agreement with reference \cite{buschlinger2015}. The lasing process was seeded via a dipole placed 10 nm along the nanowire $c$-axis away from the nanowire center, emitting a broad pulse centered at a wavelength of $\lambda_{0} = 380 \, \unit{nm}$ and having a bandwidth of $\Delta\lambda_{0} = 87 \, \unit{nm}$, corresponding to a pulse duration of $\tau_{0} = 2.4 \, \unit{fs}$. The polarization of the dipoles was chosen perpendicular to the nanowire axis in both cases. For the Ag substrate, the polarization was chosen parallel to the surface normal to the substrate, whereas for Al, it was chosen perpendicular to it, as suggested by calculations done in reference \cite{repp2023}. The corresponding longitudinal field distributions of the absolute values of the electric field for the two nanowires discussed in the main text are shown in Figure \ref{figure:NWsimul}. The top panel shows the field distributions of lasing modes for a nanowire with a diameter of $d  = 160 \, \unit{nm}$ and a length of $L = 3.6 \, \unit{\upmu m}$ on (a) Ag and (b) Al. Both field distributions overlap strongly with the nanowire. The dominating field component for both distributions is polarized out of plane, so perpendicular to the nanowire axis and the substrate plane normal vector. For the field distribution on the Ag substrate, shown in Figure \ref{figure:NWsimul}(a), there is a strong field enhancement close to the metal plane, corroborating the hypothesis that surface plasmons are excited in this configuration, resulting in the predominance of the hybrid mode. The far-field patterns shown in Figure 4 of the main text, however, exhibit a polarization oriented along the nanowire axis for this nanowire on silver. We hypothesize that this is a result of a field enhancement of the polarization component along the nanowire axis at the end facets and the nanowire itself acting as an efficient scatterer of this mode field component into the far field. This behavior is similarly observed in Figure \ref{figure:NWsimul}(b) for the Al substrate. However, since aluminum is a very good electric conductor at this wavelength, the surface plasmon is suppressed in the nanowire axis direction. Thus, the far-field pattern consists of the superposition of components parallel to the nanowire axis (i.e. a SPP-hybridized HE$_{11y}$ mode) and additional components oriented perpendicular to the nanowire axis (i.e. a SPP-hybridized HE$_{11x}$ mode), leading to the angular shift in the far-field radiation pattern.
	Figure \ref{figure:NWsimul} shows also the modal field distributions for a nanowire with a diameter of $d  = 160 \, \unit{nm}$ and a length of $L = 3.6 \, \unit{\upmu m}$ on (c) Ag and (d) Al. In this case, the lasing mode is less efficiently guided than for the thinner nanowire. Specifically, the field, which has a similar transversal structure as the HE$_{11}$ modes, travels non-collinearly to the nanowire axis and is slightly coupled out at the nanowire-air interface, leading to noticeable emission above the nanowire, which is mainly polarized orthogonal to the nanowire axis. Additionally, the surface plasmon is excited less strongly, since the effective distance between the metal and the spatial extent of the mode field is increased. On the other hand, for the Al substrate configuration, a TE$_{01}$ mode starts to lase simultaneously with the HE$_{11x}$ mode, and the resulting polarization angle is comparable to the one measured for the Ag substrate (within the estimated experimental errors).
	
	\newpage
	
	\printbibliography
	
\end{document}